\documentclass[letterpaper,notitlepage,9pt]{article}

\usepackage{kotex}
\usepackage{blindtext}
\usepackage{wrapfig} 

\usepackage{multicol}
\usepackage{graphicx}
\usepackage[superscript]{cite}
\usepackage{textcomp}
\usepackage[euler]{textgreek}
\usepackage{amsmath}
\usepackage[version=4]{mhchem}
\usepackage[dvipsnames]{xcolor}
\usepackage{datatool}

\emergencystretch=\maxdimen
\usepackage[sc]{mathpazo} 
\usepackage[T1]{fontenc} 
\usepackage{microtype} 

\usepackage[english]{babel} 

\usepackage[left=1.5cm, right=1.5cm, top=1.785cm, bottom=2.0cm]{geometry}
\usepackage[hang, small,labelfont=bf,up,textfont=it,up]{caption} 
\usepackage{abstract} 
\renewenvironment{abstract}
 {\small
  \begin{center}
  \bfseries \abstractname\vspace{-.5em}\vspace{0pt}
  \end{center}
  \list{}{
    \setlength{\leftmargin}{.5cm}%
    \setlength{\rightmargin}{\leftmargin}%
  }%
  \item\relax}
 {\endlist}

\newcommand\blfootnote[1]{%
  \begingroup
  \renewcommand\thefootnote{}\footnote{#1}%
  \addtocounter{footnote}{-1}%
  \endgroup
}

\usepackage{titlesec} 
\renewcommand\thesection{\Roman{section}} 
\renewcommand\thesubsection{\roman{subsection}} 
\titleformat{\section}[block]{\large\scshape\centering}{\thesection.}{1em}{} 
\titleformat{\subsection}[block]{\large}{\thesubsection.}{1em}{} 
\titlespacing*{\section}{0pt}{5pt plus 1pt minus .2pt}{2pt plus .2pt} 
\titlespacing*{\subsection}{0pt}{4ex plus 1ex minus .2ex}{3ex plus .2ex}

\usepackage{titling} 

\usepackage{hyperref} 

\graphicspath{\string./Images/}
\newenvironment{Figure}
  {\par\medskip\noindent\minipage{\linewidth}}
  {\endminipage\par\medskip}

\usepackage{fancyhdr}
\title{}

\begin{document}

\setlength{\headheight}{5pt}
\thispagestyle{fancy}


\begin{center}
    Using Particle Shape to Control Defects in Colloidal Crystals on Spherical Interfaces

    \vspace{0.2cm}
    Gabrielle N. Jones,\textit{$^{1}$} Philipp W.A. Sch\"onh\"ofer,\textit{$^{1}$} and Sharon C. Glotzer\textit{$^{1,2, \ast}$}
    \blfootnote{$^{\ast}$Gabrielle N. Jones (gabs@umich.edu), Philipp W.A. Sch\"onh\"ofer (pschoenh@umich.edu), and Sharon C. Glotzer (sglotzer@umich.edu)}

    \textit{\small $^{1}$~Department of Chemical Engineering, University of Michigan, Ann Arbor, Michigan 48109, USA}

    \textit{\small $^{2}$~Biointerfaces Institute, University of Michigan, Ann Arbor, Michigan 48109, USA}
\end{center}

\begin{abstract}
Spherical particles confined to a sphere surface cannot pack densely into a hexagonal lattice without defects. 
In this study, we use hard particle Monte Carlo simulations to determine the effects of continuously deformable shape anisotropy and underlying crystal lattice preference on inevitable defect structures and their distribution within colloidal assemblies of hard rounded polyhedra confined to a closed sphere surface. We demonstrate that cube particles form a simple square assembly, overcoming lattice/topology incompatibility, and maximize entropy by distributing eight three-fold defects evenly on the sphere. By varying particle shape smoothly from cubes to spheres we reveal how the distribution of defects changes from square antiprismatic to icosahedral symmetry. Congruent studies of rounded tetrahedra reveal additional varieties of characteristic defect patterns within three, four, and six-fold symmetric lattices. This work has promising implications for programmable defect generation to facilitate different vesicle buckling modes using colloidal particle emulsions.
\end{abstract}

\begin{multicols}{2}
    
\section{Introduction}
 Colloidal emulsions are found in everyday products from cosmetics to paint\cite{guzman_pickering_2022, de_carvalho-guimaraes_review_2022}. Pickering-Ramsden emulsions are a subset of this class of materials that are stabilized by the interfacial adsorption of colloidal particles rather than a molecular surfactant, such as block-copolymer stabilized polymersomes and micellar droplets\cite{ramsden_separation_1904, pickering_cxcviemulsions_1907, discher_polymersomes_2006,hunter_pickering_2020}. These colloidal emulsions have novel applications including encapsulating payloads for controlled drug delivery\cite{simovic_pickering_2011, xu_caged_2023}, facilitating interfacial bio-catalysis\cite{sun_enzymepolymer_2018}, and controlling swelling dynamics to drive particle assembly\cite{tran_swelling_2020}. They also serve as an exciting option for non-toxic and renewable emulsions with high stability, as seen in studies of cellulose particle-stabilized emulsions and foams that are shelf stable on the order of months\cite{kalashnikova_new_2011,wege_long-term_2008}. Their novel qualities arise from highly tunable properties (e.g. mechanical stability, droplet size, and porosity) that are dependent on parameters such as solvent choice\cite{qi_systematic_2014}, particle characteristics (size\cite{qi_systematic_2014}, material, shape\cite{pang_synthesis_2013, cai_synthesis_2016, yue_self-assembly_2020}, charge\cite{nallamilli_model_2014}, and surface coverage\cite{kalashnikova_new_2011}). Many of these properties are relevant for meeting cargo release\cite{sander_nanoparticle-filled_2013, li_electrostatic_2015}, retention\cite{san_miguel_permeability_2011}, and efficiency criteria\cite{huang_efficient_2024} in drug delivery applications. Examples of precise pore size control, as seen in polystyrene particle stabilized droplets, are achieved by sintering, adsorption of polymers, or particle aggregation, leading to a broad range of tunable elastic moduli and breaking forces\cite{hsu_self-assembled_2005,dinsmore_colloidosomes_2002}. 

Particle ordering in Pickering-Ramsden emulsion droplets is mirrored in many natural systems such as spherical monolayers of epithelial cells\cite{roshal_crystal-like_2020}, virus capsids\cite{klug_structure_1961,martin-bravo_minimal_2021,zandi_virus_2020,twarock_structural_2019,einert_grain_2005}, and pollen grains\cite{katifori_foldable_2010}. Models of these systems often use spherical particles constrained to an emulsion interface. The shape and pair-wise interactions of these spherical particles drive lattice structure, local symmetry, and defect morphology on the two-dimensional surface, mirroring the impact of shape and interactions in bulk crystal structure in three-dimensions\cite{guerra_freezing_2018}. On flat interfaces, spheres pack most efficiently with hexagonal ordering, where each particle $i$ has the expected coordination number of neighbors $c_i = c_{hex}=6$. Defects in hexagonal packings take on individual topological charge  $q_i = c_{hex} - c_i$, and form as isolated point disclinations or as $=\pm1$ dislocation pairs. On non-flat spaces, such as the sphere and topologically equivalent convex polyhedra, the Euler characteristic\cite{pedersen_descartes_1982} is given as $\chi = V - E + F$ where $V$, $E$, and $F$ are the number of vertices, edges, and faces, respectively, of the confining surface. Given the Euler characteristic $\chi=2$ of a spherical droplet, topology requires the preferred hexagonal packing to integrate topological defects with a total surface charge of $\sum_iq_i=12$. The required total topological charge manifests through the formation of dislocations, disclinations, or long defect ``scars'' of connected disclinations\cite{bausch_grain_2003, einert_grain_2005}. This is readily seen in the familiar patterning of a soccer ball, which has twelve five-fold $+1$ charge defects distributed with icosahedral symmetry over the surface, or in the assembly of protein subunits leading to the distinctive icosahedral buckling behavior of many virus capsids\cite{bowick_two-dimensional_2009, wang_magic_2018,clare_optimal_1991,zandi_virus_2020}. Lattice symmetry and defect structure are fundamentally related on spherical interfaces, as illustrated in molecular models of Hertzian spheres that assemble into hexagonal and simple-square lattices with competing defect motifs\cite{xie_competing_2025} and simulations of non-overlapping square tetratic molecules\cite{li_topological_2013}. 

Introducing shape anisotropy has proven to be a strong design parameter to affect assembly and packing behavior\cite{glotzer_anisotropy_2007, damasceno_predictive_2012}. Shape anisotropy affects ordering phenomena in flat space, as seen in an experimental study of superballs that showed a continuous deformation from a simple-square (cubic) to hexagonal (fcc) lattice\cite{zhang_continuous_2011, ni_phase_2012} and in numerous simulation studies of hard polyhedra,faceted spheres, and a myriad of other shapes that used shape-induced valence to drive the system towards various crystal structures\cite{haji-akbari_disordered_2009, chen_dense_2010, chen_complexity_2014, smallenburg_vacancy-stabilized_2012, avendano_phase_2012,dijkstra_m_phase_2013, teich_clusters_2016, anderson_shape_2017, shen_symmetries_2019, wang_structural_2022, sharma_effect_2023,  zhong_engineering_2024}. Studies of Pickering-Ramsden emulsions stabilized by rods\cite{rajendra_packing_2023}, cubes\cite{pang_synthesis_2013}, peanuts\cite{anjali_general_2017}, and Janus spheres\cite{luu_ellipsoidal_2014} have all reported novel ordering and demonstrated the feasibility of combining shape anisotropy and interfacial curvature. Further simulation work has shown that rods of varying asphericity adapt to spherical interfacial curvature readily, packing with nematic order with defects distributed on the vertices of a tetrahedron or about the arc of a great circle when they are orientationally locked\cite{smallenburg_close_2016, bates_nematic_2008, shin_topological_2008} and show complex two-step melting behavior when free to rotate over a range of number densities\cite{mandal_melting_2025}. 
 
In this paper, we significantly expand on these previous studies by investigating two shape families of rounded polyhedra -- cubic and tetrahedral -- constrained to a spherical surface. We focus our simulations on hard particles that form ordered, entropy-maximizing structures\cite{damasceno_predictive_2012}. We demonstrate how topology and local order conspire to dictate defect morphology and distribution about the interface. We quantify order for a variety of lattice types over a range of particle densities, sphere interface radii, and particle roundedness. Furthermore, we show the different ways in which these systems resolve the geometric frustration imposed by the incompatible topology.

\section*{Methods}
We ran hard particle Monte Carlo (HPMC) simulations of $n\in\{1000, 1500, 2000, 2500\}$ particles, with their centroid bound to the surface of a sphere with radius $R_S$. The constraining sphere surface radius $R_S$ is coupled with the particle number-area density $\rho_n=n/(4\pi R_s^2)$, as described pictorially in figure \ref{fig:shapeDefinition}a. Particle shape varies continuously between a sphere and a polyhedron (cube or tetrahedron) via a rounding parameter $s=R_{R}/(R_R+R_{PI})\in[0, 1]$, given the polyhedron insphere radius $R_{PI}$ and rounding radius $R_R$; $s = 0$ indicates no rounding of the underlying shape (sharp) and $s = 1$ indicates an isotropic sphere\cite{van_damme_classifying_2020}, as seen in figure \ref{fig:shapeDefinition}b. We clarify values of $s$ between polyhedral families by subscripts, $s_{cube}$ and $s_{tetrahedron}$. All particle volumes are normalized to the unit sphere (a sphere with radius $\sigma = 1$), $v_{particle} = \frac{4}{3}\pi$.

\begin{Figure}
  \centering
  \includegraphics[width=\textwidth]{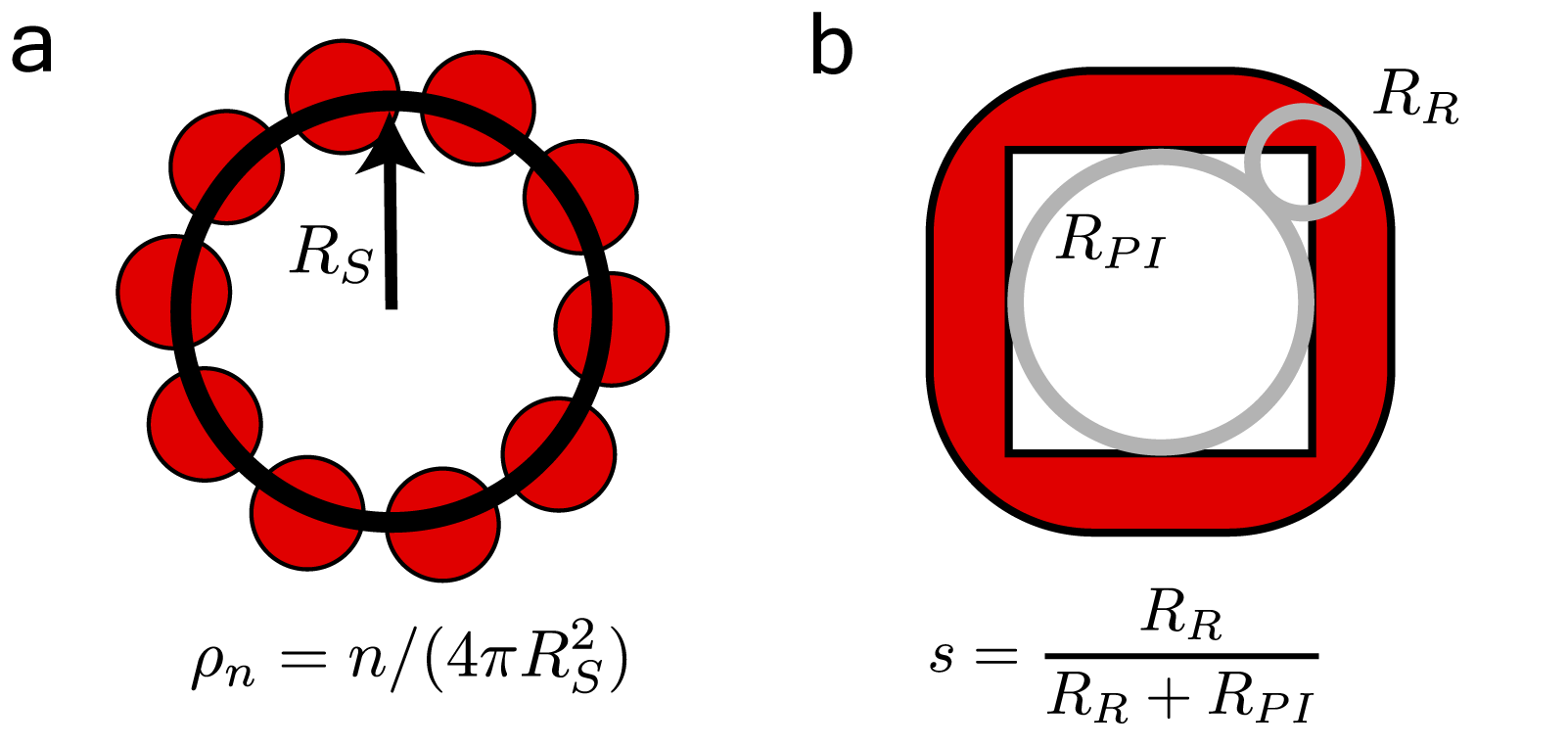}
  \captionof{figure}{Shape definition and constraining radius: (a) Particle centroids are constrained to move on a spherical interface with radius $R_S$. (b) Rounded polyhedra are defined by the union of an underlying polyhedron (cube or tetrahedron) and a rounding sphere. The rounding parameter $s$ uses the polyhedron insphere radius $R_{PI}$ and a sphere with rounding radius $R_R$ to define particle shape, seen here for a head-on view of a rounded cube.}
  \label{fig:shapeDefinition}
\end{Figure}

To obtain particle assemblies on a sphere of radius $R_S$, particles are first placed on a sphere with radius $5R_S$. The sphere is slowly reduced, over $\sim 10^6$ simulation MC sweeps, to a target radius $R_S$ and corresponding number density $\rho_n\in[0.215, 0.287]$. We simulated systems with particles in the cube family for $10^7$ MC trial moves, while tetrahedral particle systems were run for up to $4*10^7$ trial moves, with an acceptance ratio of 33\%. We found that these simulation lengths were sufficient to equilibrate the particle positions and orientations, by observing autocorrelation functions decaying to zero for the relevant system order parameters for the total number of ``ordered'' particles over the simulation length. The breadth of system parameters studied was chosen to demonstrate the robustness of our analysis methods and the general effect on the resulting assemblies of imposed topological constraints. Each combination of these parameters was simulated with 10 replicates each for $n=1000$ and 5 replicates for $n>1000$.

All simulations were performed with an add-on to the HOOMD-blue simulation toolkit (version 3.10.0)\cite{anderson_hoomd-blue_2020} that replaces translational moves in 3D Euclidean space with translational moves in $S^2$. This modified HPMC algorithm is modeled after a schema described for simulations in hyperspherical geometry\cite{sinkovits_nanoparticle-controlled_2012,schonhofer_rationalizing_2023} and provides a robust and efficient particle displacement method. Particle translation and rotation are uncoupled, allowing for full three-dimensional rotational degrees of freedom of the particles. All particles interact purely via excluded volume interactions. We used the freud\cite{ramasubramani_freud_2020} python library to analyze our simulations and the signac\cite{adorf_simple_2018} software package for data management. Simulation snapshots are visualized using Ovito visualization software\cite{stukowski_visualization_2010}.

All order parameters are normalized such that a value of $1$ denotes perfect agreement with the desired order and $0$ denotes low agreement. Although each order parameter is normalized to the same range, they use parameter specific cutoffs to denote particles within their respective lattice motif. These cutoffs are informed by probability distributions. Detailed explanation of order parameter calculations and justifications for cutoffs are included in the supplementary information section S2. The variety of per-particle order parameters described are averaged over the first nearest-neighbor shell of the particle using a distance cutoff to give a subscript $nn$. This cutoff was calculated to be between the first two peaks of the geodesic radial distribution function $g(\theta)$. We use four different order parameters to describe the four crystalline motifs found in this study:

    1. \textit{The hexagonal motif:} The hexagonal order parameter, $|\psi_6|_{nn}$, quantifies local six-fold orientational symmetry via a per-particle value of $\psi_6=\frac{1}{n}\Sigma^n_je^{6i\theta_{ij}}$. We assert that hexagonal packing describes particles with $|\psi_6|_{nn} \ge 0.80$ and consider particles with $|\psi_6|_{nn} < 0.80$ to be defects. This cutoff is in agreement with previous studies of hard-sphere-like colloidal particles of silica and polymethylmethacrylate\cite{wu_melting_2009}. 

    2. \textit{The face-aligned motif: } We quantify the face-to-face alignment of rounded cubes by the order parameter $f_{nn}$. This quantifies how parallel the closest faces of two neighboring cubes are, rotated to be in the same tangential reference plane. We use a cutoff of $f_{nn} \ge 0.75$ to describe highly face-aligned rounded cubes, and consider particles with $f_{nn}< 0.75$ to be defects.

    3. \textit{The honeycomb motif: } The honeycomb order parameter, $hc_{nn}$, captures the high local three-fold orientational symmetry of neighboring tetrahedra. This motif has a restriction that particles must be vertex-up or face-up, corresponding to the basis positions of the honeycomb point set. We use a cutoff of $hc_{nn} \ge 0.75$ to describe the honeycomb motif, and consider particles with $hc_{nn}< 0.75$ to be defects. This motif is similar to the honeycomb phase of tetrahedron superlattices on substrates\cite{zhou_chiral_2022}.

    4. \textit{The woven motif: } The woven order parameter, $w_{nn}$, for rounded tetrahedra is described using a two-particle arrangement of ``edge up'' tetrahedra rotated by $\pi/2$ relative to each other. This order parameter captures the motif's distinct ``interwoven'' appearance and tetratic symmetry. We use a cutoff of $w_{nn} \ge 0.65$ to describe the woven motif, and consider particles with $w_{nn}< 0.65$ to be defects. 

\textit{Description of defect morphology. } We expand the concept of defect scars to examine the shape of the defects for spheres with $s = 1$, cubes with $s_{cube} = 0$, and tetrahedra with $s_{tetrahedron} = 0$. We binarize the systems using $|\psi_6|_{nn} < 0.80$, $f_{t, nn} < 0.75$, and $w_{nn} < 0.65$ for the sphere, cube, and tetrahedron systems, respectively. We do not consider defect morphologies of honeycomb dominated systems due to a lack of isolated defect regions. We construct a connected graph with these defect particles as nodes and connect these nodes based on a distance cutoff corresponding to the first neighbor peak within the geodesic radial distribution function. For each component of the graph (isolated defect) the defect length $d[\sigma]$ is defined as the longest shortest path. We study defect region length as the system size, $n\in{1000, 1500, 2000, 2500}$, increases over a range of $\rho_n$.

\section*{Results and Discussion}
\begin{figure*}
  \centering
  \includegraphics[width=\linewidth]{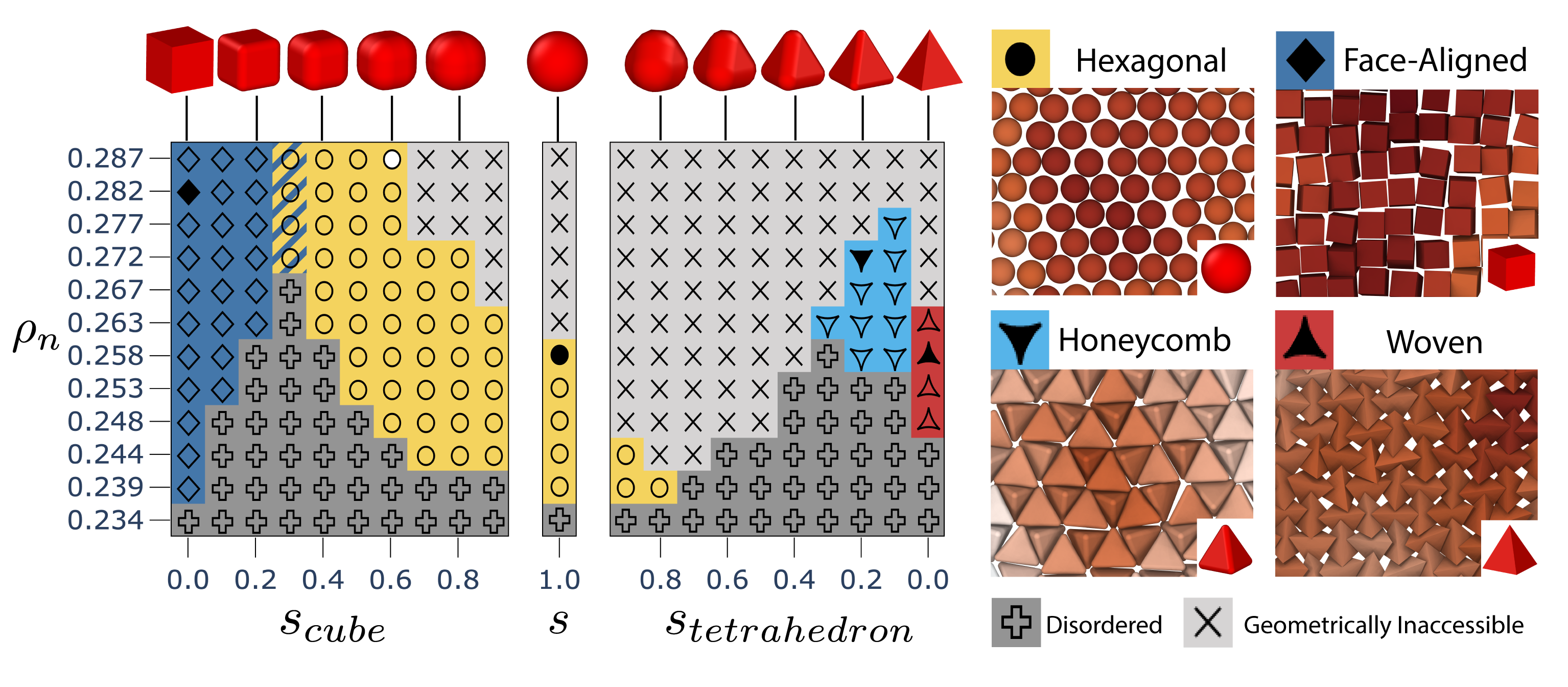}
  \caption{\textbf{Phase diagrams for systems with $n=1000$ particles over a range of particle roundedness $s$ and number densities $\rho_n$.} For cubic rounded polyhedra the face-aligned motif dominates at low rounding values $s_{cube}$, gradually shearing as the rounding of the shape increases, and exhibits high hexagonal ordering as $s_{cube}$ approaches 1. Striped regions at $s_{cube} = 0.3$ and $\rho_N\ge0.272$ indicate regions where both hexagonal and face-aligned motifs are observed. When $s = 1$ (middle), idealized spheres assemble the hexagonal motif. At high values of $s_{tetrahedron}$ the hexagonal motif continues to dominate. With intermediate rounding, tetrahedra assemble the honeycomb structure before giving way to the woven structure as $s \rightarrow 0$. The four distinct motifs (right) are color coded to match the phase diagram, with motif insets corresponding to the appropriate black-filled marker. These specific motifs correspond to the following values of (shape, $s$, $\rho_n$, marker): (Sphere, $1.0$, $0.258$, yellow circle), (Cube, $0.0$, $0.282$, dark blue diamond), (Tetrahedron, $0.2$, $0.272$, light blue downward triangle), (Tetrahedron, $0.0$, $0.258$, red upturned triangle) from left to right, top to bottom. The white filled marker at $s_{cube} = 0.6$ and $\rho_n = 0.287$ indicates a jammed state rather than self-assembled structure.}
  \label{fig:phaseMotif}
\end{figure*}

We constructed phase diagrams for assemblies of $n = 1000$ particles for our two shape families of rounded polyhedra as shown in figure \ref{fig:phaseMotif}. The various phases reported in the figure demonstrate our ability to change the self-assembled crystal structure by varying particle shape and roundedness $s$. 

\begin{figure*}
  \centering
  \includegraphics[width=0.5\textwidth]{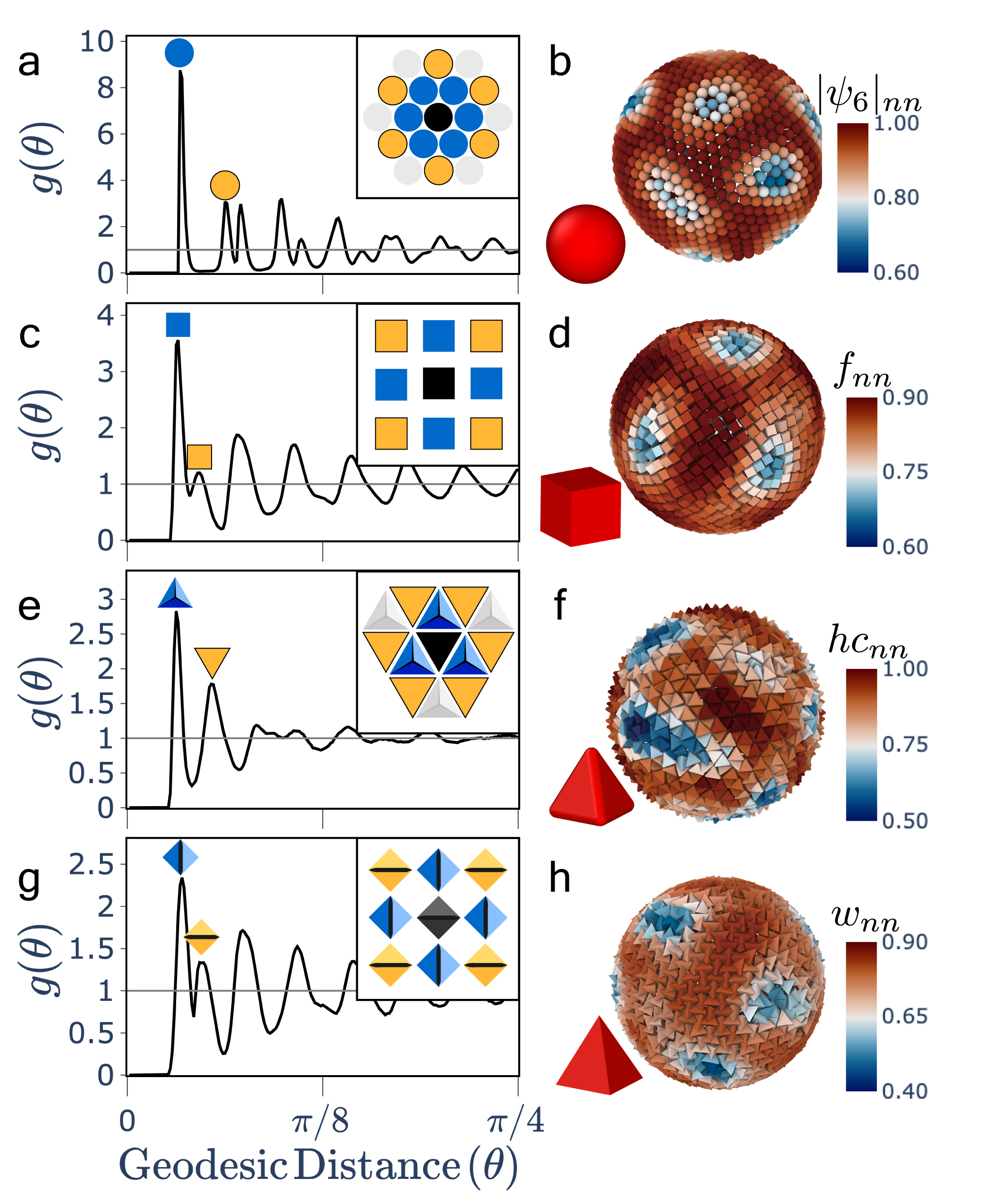}
  \caption{Reference radial distribution function $g(\theta)$ plots for each motif: (a) A representative $g(\theta)$ for the hexagonal motif, taken from simulations of spheres with $s=1$ and $\rho_n = 0.258$. (b) A simulation image at these parameters colored by $|\psi_6|_{nn}$. (c) Representative $g(\theta)$ for the face-aligned motif, most commonly realized in a simple-square lattice, taken from simulations of rounded cubes with $s=0$ and $\rho_n = 0.282$. (d) A simulation image at these parameters colored by $f_{nn}$. (e) Representative $g(\theta)$ for the honeycomb motif, taken from simulations of rounded tetrahedra with $s=0.2$ and $\rho_n = 0.267$.  (f) A simulation image at these parameters colored by $hc_{nn}$. (g) Representative $g(\theta)$ for the woven motif, taken from simulations of rounded tetrahedra with $s=0$ and $\rho_n = 0.263$. (h) A simulation image at these parameters colored by $w_{nn}$.} 
  \label{fig:rdfReference}
\end{figure*}

\begin{figure*}
  \centering
  \includegraphics[width=\linewidth]{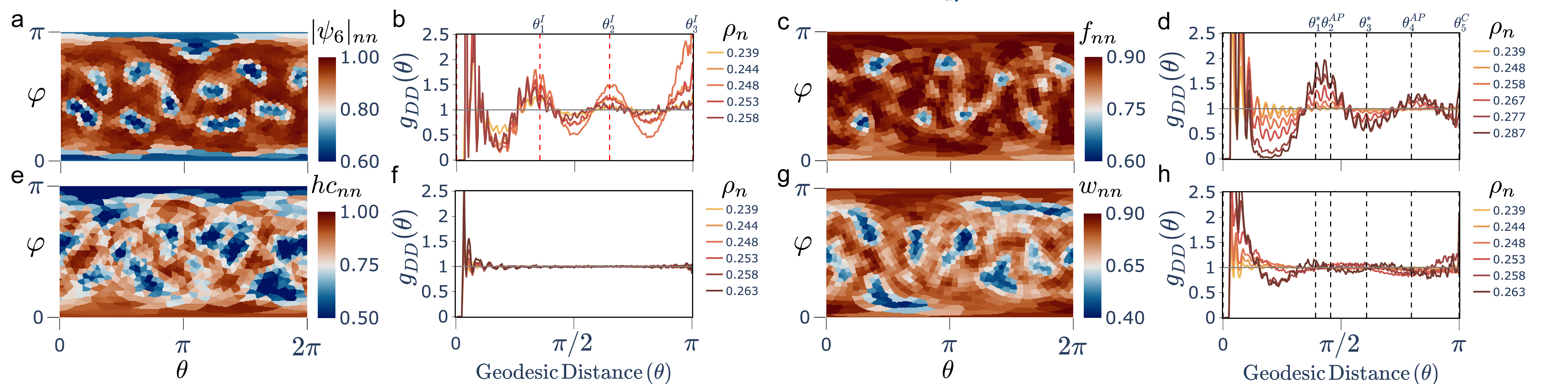}
  \caption{Defect Distribution: Defect distribution is presented in Mercator projections, and quantified by the geodesic radial distribution of defects, $g_{DD}(\theta)$, across number densities $\rho_n$. Systems are binarized by a value noted by the midpoint of their respective color bars seen in the third column. (a) In a system of $n=1000$ spheres there are twelve isolated defect regions. (b) The distribution of these defects is icosahedral, seen by peaks in $g_{DD}(\theta)$ at geodesic distances corresponding to the vertices of an icosahedron. The representative system has a density of $\rho_n = 0.258$. (c) In systems of $n=1000$ cubes, with a value of $s=0$, there are eight isolated defects. (d) Their distribution is square antiprismatic, as seen in peaks in $g_{DD}(\theta)$ at $\theta_1, \theta_2^{AP}, \textrm{ and }\theta_4^{AP}$. There are no peaks at $\theta_3$ and $\theta_5^{C}$ that would indicate a cubic distribution. The representative system has a density of $\rho_n = 0.282$. (e) In systems of $n=1000$ rounded tetrahedra with $s=0.2$, there is no regular distribution of defects. (f) Given this disorder the defects show no correlation or coordination across the surface, as $g_{DD}(\theta)$ remains small across all values of $\rho_n$. The representative system has a density of $\rho_n = 0.267$. (g) In systems of $n=1000$ tetrahedra, with a value of $s=0$, eight isolated defects appear only at a high density $\rho=0.263$. (h) The spatial distribution of defects shows little to no correlation across the surface for all densities $\rho_n< 0.263$, with the first trough in $g_{DD}(\theta)$ indicating only that defect regions are spatially isolated from each other. Each of the Mercator projections shown here (a, c, e, g), correspond to the simulation images shown in figure \ref{fig:rdfReference}(b, d, f, h).}
  \label{fig:DefectDistribution}
\end{figure*}

\begin{figure*}
  \centering
  \includegraphics[width=\linewidth]{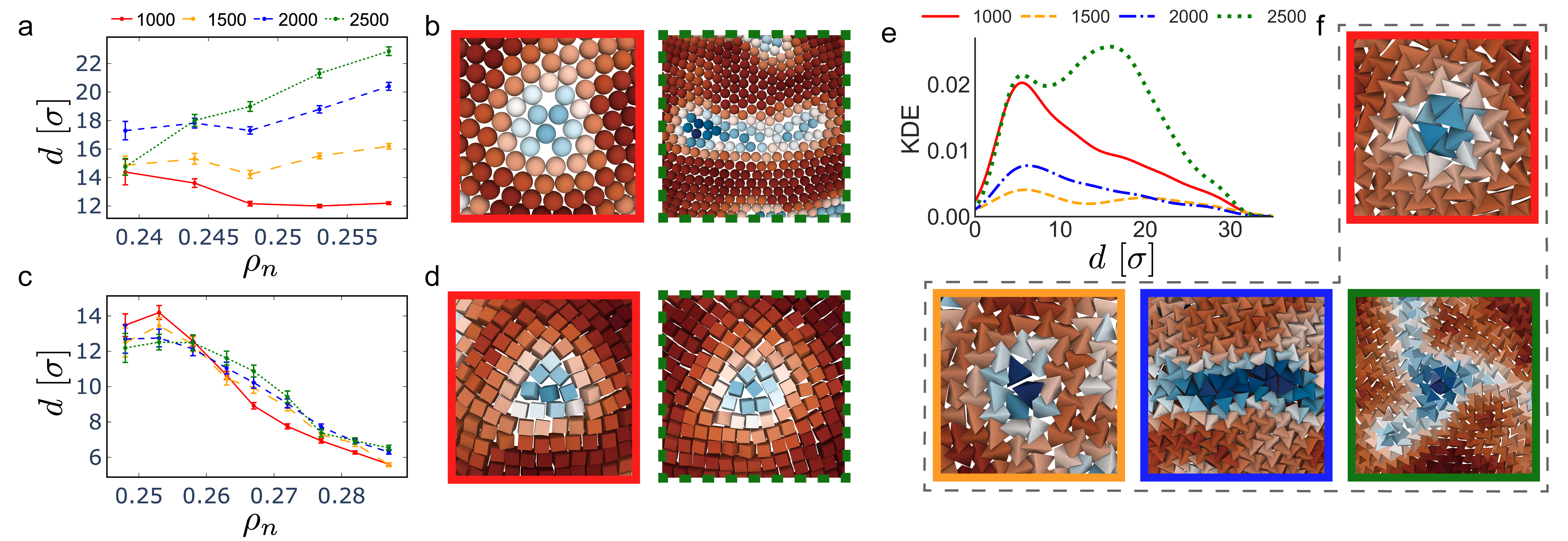}
  \caption{Defect lengths $d[\sigma]$ are measured as the longest shortest path in the simply connected graph formed by defect particles for each system over a range of particle number $n = {1000, 1500, 2000, 2500}$. (a) Defect lengths as a function of $\rho_n$ are shown for spheres with $s=1$, generally increasing as a function of $n$ for sufficiently high $\rho_n$. (b) This increase in $d[\sigma]$ is shown in representative defect images at $n = 1000 \textrm{ (left) and } 2500 \textrm{ (right)}$, at a density $\rho_n = 0.258$. (c) In systems of rounded cubes with $s_{cube}=0$, defect lengths can be seen to generally decrease as $\rho_n$ increases, across all values of $n$. (d) The length $d[\sigma]$ is consistent across $n$, with $n = 1000 \textrm{ (left) and } 2500 \textrm{ (right)}$, seen in representative defects with $\rho_n = 0.287$. (e) Rounded tetrahedra with $s_{tetrahedron}=0$ form a large variety of defects, with broadly distributed lengths, seen for $\rho_n = 0.263$. (f) A range of exemplary defects are shown over the range $n = {1000, 1500, 2000, 2500}$, top to bottom, left to right.}
  \label{fig:scarLengths}
\end{figure*}

\textit{Results for spheres: } The phase diagram in figure \ref{fig:phaseMotif} reveals that the dominant lattice order for assembles of $n=1000$ spheres is hexagonal across a range of number densities $\rho_n \ge 0.239$, with a simulation image in figure\ref{fig:DefectDistribution}a. The ordering of the hexagonal motif is supported by distinct peaks in the radial distribution  function $g(\theta)$ and seen in a simulation snapshot in figures \ref{fig:rdfReference}a,b. The hexagonal lattice is disrupted by defect scars that each have the expected $+1$ overall topological charge, and for sufficient $\rho_n$ we identify 12 distinct scars with low $|\psi_6|_{nn}$ in an unwrapped Mercator projection of the system in figure \ref{fig:DefectDistribution}a. These 12 distinct defect regions are distributed with icosahedral symmetry, as seen in figure \ref{fig:DefectDistribution}b. Peaks in the geodesic radial distribution function calculated between defects, $g_{DD}(\theta)$, appear at geodesic distances $\theta_{1}^I, \theta_{2}^I, \textrm{and } \theta_{3}^I$. These distances are the geodesic distances between 12 points distributed on a sphere with icosahedral symmetry. As we increase system size by increasing $n$ at constant $\rho_n$ the defect scar length $d[\sigma]$ increases as seen in figures \ref{fig:scarLengths}a,b. This is in agreement with previous studies of spheres on the surface of a sphere\cite{guerra_freezing_2018} and serves as a baseline for our studies of cubic and tetrahedral rounded polyhedra.

 \textit{Results for rounded cubes:} We investigate the family of rounded cubes by decreasing $s_\text{cube}$, with constant $n=1000$. For highly rounded cubes with $s_\text{cube}\ge0.4$ a hexatic rotator phase dominates, similar to experimental and simulated systems of rounded cubes\cite{zhong_engineering_2024} on flat substrates and superballs\cite{ni_phase_2012} in bulk. This phase has high hexatic order for $\rho_n\ge 0.244$ and little orientational correlation between particles, as indicated by the high and low global averages of $\langle|\psi_6|_{nn}\rangle$ and $\langle f_{nn}\rangle$, respectively. The crystallization transition gradually increases to a maximum value $\rho_n=0.272$ at $s_\text{cube}=0.3$. At this rounding value the particles form both hexagonal and face-aligned motifs, indicating a coexistence of spatially disparate domains within one system. As the particle shape approaches a sharp cube with $s_\text{cube} = 0$, face-aligned order becomes dominant and gives rise to a simple square lattice with tetratic symmetry and high orientational order, shown in figures \ref{fig:phaseMotif} and \ref{fig:rdfReference}c. An example of this phase is seen in figure \ref{fig:rdfReference}d. Furthermore, the onset number density for crystallization rapidly decreases to $\rho_n=0.239$ at $s_\text{cube} = 0$.
 
 From the simulation image in figure \ref{fig:DefectDistribution}d and its Mercator projection in figure \ref{fig:DefectDistribution}c, we see isolated defect regions colored by low values of $f_{nn}$. The preferred local symmetry around particles changes from hexagonal to tetragonal leading to an expected coordination number of $c_i = c_{square} = 4$. From the topological expectations that arise from the Euler characteristic $\chi = 2$ and the dominant tetragonal symmetry, we rationalize the symmetric distribution of eight three-fold symmetric $q=+1$ topological charges, figure \ref{fig:DefectDistribution}c, similar to $k=-1/2$ defects seen in nematic liquid crystals\cite{tang_orientation_2017}. Spatial correlations between face-aligned defect particles as quantified by $g_{DD}(\theta)$ in figure \ref{fig:DefectDistribution}d show an organized distribution of defects about the surface. We hypothesize that there are two candidates for the distributed symmetry of these defects, one aligning to the vertices of a cube and the other to a square antiprism. Geodesic distances for these symmetries are denoted by $\theta_i$, $i\in[1, 2, 3, 4, 5]$, and the superscript denotes whether this distance is attributed to cubic (C) symmetry, square antiprismatic (AP) symmetry, or both (*). Peaks in $g_{DD}(\theta)$ show high correlations corresponding to $\theta_1^*, \theta_2^{AP}, \textrm{ and } \theta_4^{AP}$, demonstrating that the distribution aligns more closely with the symmetry of the square antiprism, reminiscent of a study of hard squares on the sphere by Li et. al\cite{li_topological_2013}. This distribution of defects has more space between defects and is the solution to the Thompson problem for eight repulsive points distributed about a sphere surface. 
 
 The three-fold nature of the defect region is showcased in figure \ref{fig:scarLengths}d, appearing through a cooperative loss of orientational order and attributed to topological resolutions to lattice incompatibility. 
 The defect has either one or three particles centered about the geometric center of the defect; leading to the two peaks in $d[\sigma]$ shown in the supplementary, figure S13. These central particles help stabilize the defect and rotate more freely compared to particles on the periphery or outside of the defect, rather than forming as a result of jamming between nearby misoriented grains. As particle number $n$ increases, the absolute length of the defect $d[\sigma]$ remains constant, but decreases with increasing $\rho_n$ as seen in figures \ref{fig:scarLengths}c, d.
 
 We observed no smooth transition from five-fold defect scars to localized three-fold symmetric defects with corresponding icosahedral and square antiprismatic defect distribution symmetry as $s_{cube}$ decreases. Instead, we observed a dissolution of one defect morphology and a subsequent emergence of the counter morphology. The dominant translational order transition between hexagonal and simple-square is different to those of superball assemblies in bulk\cite{zhang_continuous_2011} where lattice transformation occurs via a smooth variable angle shear square transition as the roundedness of the particles decreases. Confined to the surface of a sphere, bounded system size and required topological defects hinder the system's ability to smoothly interpolate between these two ends of the spectrum. We also note the monotonic increase in allowable $\rho_n$ as $s_{cube}$ decreases, which follows from the high maximum number density $\rho_n = 0.385$ given in the flat $R=\infty$ limit with defect-free simple-square packing. 
 
 \textit{Results for rounded tetrahedra: } We repeated the simulations with the family of rounded tetrahedra by starting from sphere particle systems of size $n=1000$ and decreasing $s_\text{tetrahedron}$.  Rounded tetrahedra show very distinct phase regions as a function of $s_{tetrahedron}$; we observed no gradual decrease of one order parameter corresponding to an increase in another. This leads to a phase diagram with large regions that are geometrically inaccessible for rounded tetrahedra but accessible to rounded cubes.

 As we begin to decrease $s_{tetrahedron}$ the hexatic rotator phase quickly gives way to a disordered phase. This rotator phase is disallowed at high $\rho_n$ and moderate $s_{tetrahedron}$ due to the effective radius of the particle as it rotates. The circumsphere radius of rounded tetrahedra grows much more quickly for tetrahedra as compared to cubes, and correspondingly the maximally allowed $\rho_n$ for perfect hexagonal packing sharply decreases with decreasing $s_{tetrahedron}$. Decreasing $s_{tetrahedron}$ further, we see the emergence of a honeycomb phase. The first neighbor shell includes three particles that align face to face with the center particle, and are at distances larger than the minimal steric distance. For example, given $\rho_n = 0.263$ and $s_{tetrahedron} = 0.3$, the average Euclidean distance to the first nearest neighbors is $1.754$, compared to a minimal Euclidean distance of $1.434$. This phase does not span the surface at any simulated $\rho_n$, as evidenced by the lack of long range correlations in $g(\theta)$, figure \ref{fig:rdfReference}e. The system in figures \ref{fig:rdfReference}f and \ref{fig:DefectDistribution}e has exceptional honeycomb coverage compared to the majority of replicate simulations. Correspondingly, the assembled structures lack consistent and clearly isolated defect regions, with no spatial correlations in their distribution, figure \ref{fig:DefectDistribution}f.

 As $s_{tetrahedron} \textrm{ approaches } 0$ the woven motif appears for a small range of number densities $0.248 \le \rho_n \le 0.263$. This motif has tetratic symmetry and forms a simple square lattice as evidenced by the peaks in figure \ref{fig:rdfReference}g, bearing a marked similarity to the simple square lattice of face-aligned cubes in figure \ref{fig:rdfReference}c. When rounding is minimally varied within the region $0<s_{tetrahedron}<0.1$ (see supplementary S2, figure S10) we see that the woven motif is highly sensitive to particle rounding and persists for a small window of $s_{tetrahedron} < 0.06$ before the honeycomb motif returns. For $\rho_n > 0.263$ and $s_{tetrahedron} = 0$ the surface area constraint prevents further assembly. The woven motif forms a tightly bound network, maintaining favorable face to face alignment with four neighbors. The minimum distance between particles in this motif with $s_{tetrahedron} = 0$ would be $1.644\sigma$. Yet, at $s_{tetrahedron}$ and $\rho_n = 0.263$ the average distance to the first neighbor shell is $1.914\sigma$, larger than the minimum distance due to warping of the woven motif due to the surface curvature.

 For sharp tetrahedra with $s_{tetrahedron} = 0$, a single, surface-spanning woven cluster forms only at a density of $\rho_n = 0.263$. Although the woven motif has tetratic symmetry, defects rarely exhibit the topologically expected three-fold symmetric morphology and even more rarely form eight ideally isolated defects. The instance of clearly isolated defects given in figure \ref{fig:DefectDistribution}g is exceptional. In spite of this, peaks in $g_{DD}(\theta)$ show that defects in systems of woven tetrahedra display little to no long range correlations, figure \ref{fig:DefectDistribution}h. The defect-ridden assembly of the woven phase is attributed to the lack of reconfigurability of the lattice. In comparison to the face-aligned simple square lattice seen in cubes, the woven motif is translationally and orientationally rigid in its construction; this prevents the resolution of misalignment between two grains of woven tetrahedra by sliding. As the number of particles $n$ increases the defects in ordering continue to lack a clear morphology, and the distribution of defect lengths $d[\sigma]$ remains wide, figure \ref{fig:scarLengths}e. Some of these defects are highly localized, but a variety of morphologies are present, including elongated scars and three-fold chiral spirals, shown in figure \ref{fig:scarLengths}f.

\section*{Conclusions}

 We have presented work that seeks to provide insight into the effect of spherical topology on the self-assembled ordering of rounded polyhedra and the distribution of topological defects. Hard spheres exhibit hexagonal order, and distribute defects with icosahedral symmetry. By continuously changing the particle shape to cubes, the entropic preference for face-to-face alignment drives the system towards simple square order, with a notable lack of a smooth shear-square transition. Nonetheless, the topological incompatibility of the local tetragonal symmetry seen in cubic rounded polyhedra and the surface drives the formation of eight three-fold symmetric isolated defects, distributed with approximately square antiprismatic symmetry about the surface. For our other shape family, as we morph spherical particles into tetrahedra we see the appearance of the honeycomb and woven motifs. These self-assembled motifs are highly dependent on utilizing the full three-dimensional rotational degrees of freedom of the tetrahedron. For $n=1000$ particles, both motifs give rise to incoherent and non-distinct defect regions. As we increase $n$ for sharp tetrahedra we continue to see a wide distribution of defect morphologies, only rarely observing distinctly three-fold symmetric defects. Overall, the reconfigurability of the assembled lattice impacts how evenly topological strain is localized and distributed. 
 
 In systems of anisotropic particles at interfaces, entropic interactions are only one of many important contributions. Even particle orientation and depth of the particle centroid within the interface become functions of particle wettability and fluid-fluid surface tension\cite{botto_capillary_2012,anjali_contact_2016, kumar_phase_2021}. Theoretical studies of cubic particle assemblies at flat fluid-fluid interfaces have predicted orientations relative to the interface normal (face, edge, or vertex up) with varying frequency\cite{anzivino_chains_2021}. Interparticle interactions dictated by orientationally-dependent induced capillary forces between particles can lead to a variety of non-trivial assemblies, as observed in experimental systems of cuboidal hematite particles\cite{anjali_shape-induced_2017}.  As the separation between defects drives buckling transitions in icosahedral virus capsids, our observation of eight consistent three-fold defects opens avenues for new buckling phenomena. Vesicles with a faceted cuboid morphology have been observed in phospholipid liposomes by enhancing intra-membrane attraction forces to overcome surface tension, yet there is variation in aspect ratio of these vesicles\cite{bakardzhiev_unprecedented_2021}. Furthermore, complicated buckling modes have been shown to appear in simulated and natural systems, such as the infolding of desiccated pollen grains, preventing vesicle collapse and allowing for later re-hydration\cite{yuan_crystalline_2019, vernizzi_platonic_2011, bozic_mechanics_2022, matoz-fernandez_pymembrane_2023, waltmann_patterning_2024}. 
 
 We have shown that shape anisotropy is a strong design parameter to influence defect morphology and distribution. Our findings within this minimal entropic model show that defect distribution varies in non-trivial ways and the morphologies that have been shown in the paper are unlike those in systems of spherical particles. The total number of isolated defects and their characteristic topological charge vary as a function of particle number, intrinsic shape, and particle rounding, showing that these tunable parameters dictate particle order and defect distribution and separation. By showing the self-assembly behavior of crystal motifs and the distribution of defects in hard particle systems to be designable and robust we hope to inspire new directions in colloidosome design. 

\section{Author contributions}

GNJ and PWAS were both heavily involved in the conceptualization of the study and software development. GNJ performed the simulations and analyzed the data. All authors contributed to the investigation and the writing of the manuscript. SCG directed the research.

\section{Conflicts of interest}
There are no conflicts to declare.

\section{Acknowledgments}

We would like to acknowledge interesting discussions on this work with Prof. Jonathan Selinger, Prof. Xiaoming Mao, and Dr. Nan Cheng. 

Research supported in part by the Center for Complex Particle Systems (COMPASS) a National Science Foundation Science and Technology Center, Division of Materials Research Award \#DMR 2243104. GNJ acknowledges support from the U.S. Department of Energy, Office of Science, Office of Advanced Scientific Computing Research, Department of Energy Computational Science Graduate Fellowship under Award Number DE-SC0022158.  This work used Anvil at Purdue RCAC through allocation DMR 140129 from the Advanced Cyberinfrastructure Coordination Ecosystem: Services \& Support (ACCESS) program, which is supported by National Science Foundation grants \#2138259, \#2138286, \#2138307, \#2137603, and \#2138296.  Computational resources and services were also provided by Advanced Research Computing (ARC), a division of Information and Technology Services (ITS) at the University of Michigan, Ann Arbor.

Disclaimer: "This report was prepared as an account of work sponsored by an agency of the United States Government. Neither the United States Government nor any agency thereof, nor any of their employees, makes any warranty, express or implied, or assumes any legal liability or responsibility for the accuracy, completeness, or usefulness of any information, apparatus, product, or process disclosed, or represents that its use would not infringe privately owned rights. Reference herein to any specific commercial product, process, or service by trade name, trademark, manufacturer, or otherwise does not necessarily constitute or imply its endorsement, recommendation, or favoring by the United States Government or any agency thereof. The views and opinions of authors expressed herein do not necessarily state or reflect those of the United States Government or any agency thereof."

\bibliography{Michigan} 
\bibliographystyle{ieeetr}

\end{multicols}

\end{document}


\onecolumn

\tableofcontents

\newpage

\section{S1: Autocorrelation Decay of Order Parameters}

\begin{figure}[h!]
  \centering
  \includegraphics[width=\linewidth]{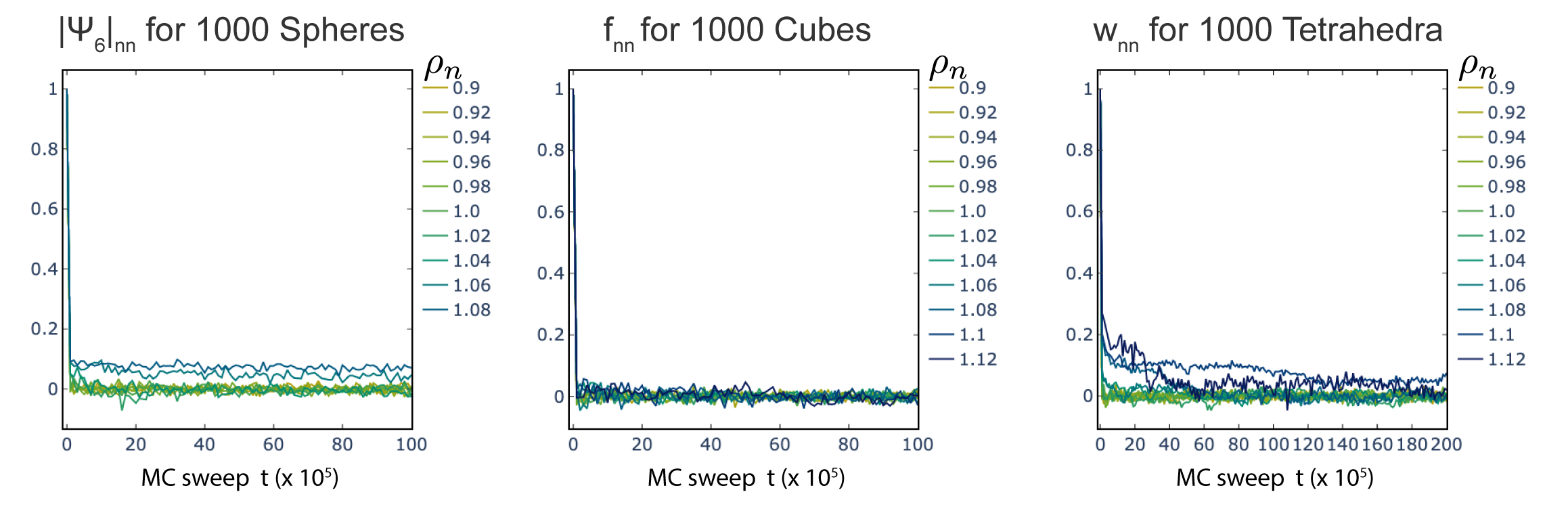}
  \caption{Autocorrelation of per-particle $|\psi_6|_{nn}$, $f_{nn}$, and $w_{nn}$ order parameters for spheres ($s = 1$), cubes ($s_{cube}=0$), and tetrahedra ($s_{tetrahedron}=0$). Some simulations of tetrahedra were run for up to $4*10^7$ timesteps to meet equilibrium criteria, which are not shown here for simplicity. Color corresponds to number density $\rho_n$.}
  \label{fig:Autos}
\end{figure}


\section{S2: Description of Order Parameters}
 Local order parameters are widely used to quantify how close particles are to a target crystalline motif. Cutoff criteria to determine crystallinity are often stated pragmatically, but this section will clearly lay out each specific analyses, what they capture, and justification for each order parameter cutoff.

\begin{figure}[h]
  \centering
  \includegraphics[width=\linewidth]{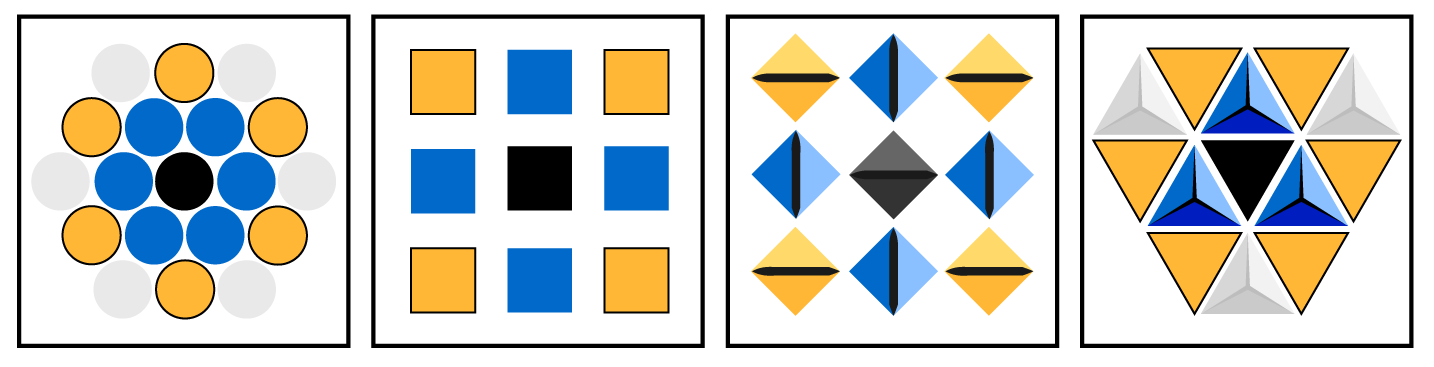}
  \caption{Illustration of the four dominant motifs. From left to right they are the hexagonal, face aligned, woven, and honeycomb motifs. They are described by $|\psi_6|$, $f$, $w$, and $hc$, respectively.}
  \label{fig:MotifExamples}
\end{figure}

\subsection{Defining rotational symmetry}

\begin{figure}[h]
  \centering
  \includegraphics[width=\linewidth]{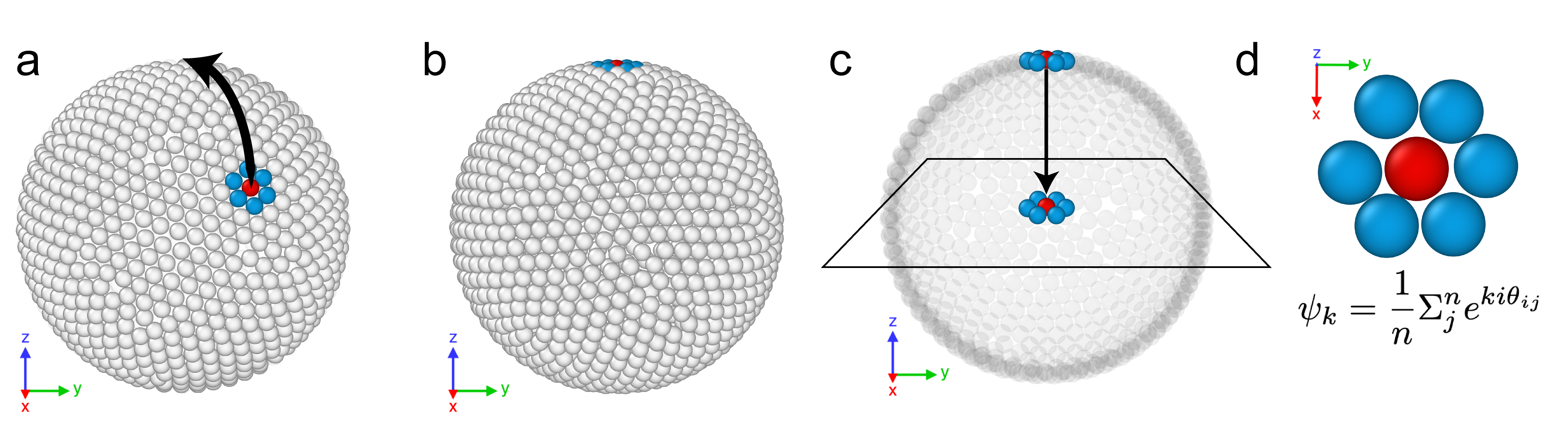}
  \caption{Calculation of $\psi_k$ on the surface of a sphere. (a) A particle $i$ (red) is chosen, and $k=6$, in this case, neighbors are chosen (blue). (b) They are then rotated to the $+z$ pole, positioned at $(0, 0, R_S)$. (c) The relevant particles are projected onto the xy plane, preserving interparticle angles but not distances. (d) Now, with a top-down view, the k-atic order parameter can be calculated.}
  \label{fig:PsiKOnSphere}
\end{figure}

The general k-atic order parameter is given by $\psi_k=\frac{1}{n}\Sigma^n_je^{ki\theta_{ij}}$; where j denotes elements of the set of the $k$ nearest neighbors. For this calculation all particles on the surface are rotated such that particle $i$ is positioned at $(0, 0, R_S)$, and all particles $j$ are projected from the surface onto the tangential plane at particle $i$, the xy plane. Following this projection, $\vec{r}_{ij}$ is the center to center vector from particle $i$ to particle $j$, and $\theta_{ij}$ is the angle between the vector $(1, 0)$ and $\vec{r}_{ij}$. This projection-calculation scheme for the k-atic order parameter preserves $\theta_{ij}$ but not the true geodesic distance between particles $\vec{r}_{ij}$. An illustration of this method can be seen in figure \ref{fig:PsiKOnSphere}. 

\subsection{Defining orientation}

\begin{figure}[h]
  \centering
  \includegraphics[width=\linewidth]{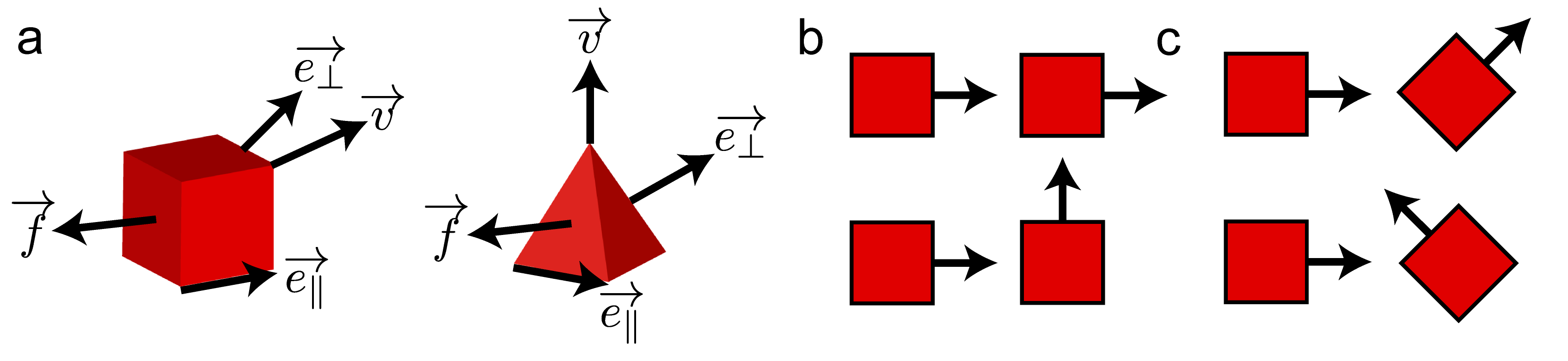}
  \caption{(a) Illustration of the four sets of associated shape vectors that we use in this paper, shown for the cube (left) and tetrahedron (right) shape families. Illustration of equivalent rotations. (b) This shows orientations for the class of face normal vectors that equate to a value of $\theta_{ij} = 0$ (top) and $\theta_{ij} = \pi/2$ (bottom). (c) This shows orientations in the same class that equate to a value of $\theta_{ij} = \pi/4$ (top) and $\theta_{ij} = 3\pi/4$ (bottom).}
  \label{fig:VectorAndAngleExample}
\end{figure}

For each non-sphere shape ($s\ne 1$) there are four sets of associated vectors, as seen in figure \ref{fig:VectorAndAngleExample}a.
\begin{itemize}
  \item Face Normals ($\overrightarrow{f}\in \{F\}$)
  \item Vertices ($\overrightarrow{v} \in \{V\}$)
  \item Edge Normals ($\overrightarrow{e_\perp} \in \{E_\perp\}$)
  \item Edge Tangents ($\overrightarrow{e_\parallel} \in \{E_\parallel\}$)
\end{itemize}

For a given set of vectors within each class $\overrightarrow{a}_i \in A$ and $A \in \{F,V, E_\perp, E_\parallel\}$, we find the vectors that minimize the angle $\theta$ between both the surface normal vector ($\theta_n$) and the surface tangent plane ($\theta_t$). For each class we can calculate the following values,

\begin{center}
\begin{tabular}{ c|c|c }
 & minimal angle & associated minimal angle vector \\
 \hline
 relative to surface normal & $\theta_{n,A}$ & $\overrightarrow{a},n$ \\
 relative to surface tangent & $\theta_{t,A}$ & $\overrightarrow{a},t$ \\
\end{tabular}
\end{center}

We classify particles as face, vertex, or edge-up by the class corresponding to the minimal $\theta$ value out of the set $\{\theta_{n, F}, \theta_{n, V}, \theta_{n, E_\perp}\}$.

\subsection{Description of the hexagonal motif}
\begin{figure}[h!]
  \centering
  \includegraphics[width=0.8\linewidth]{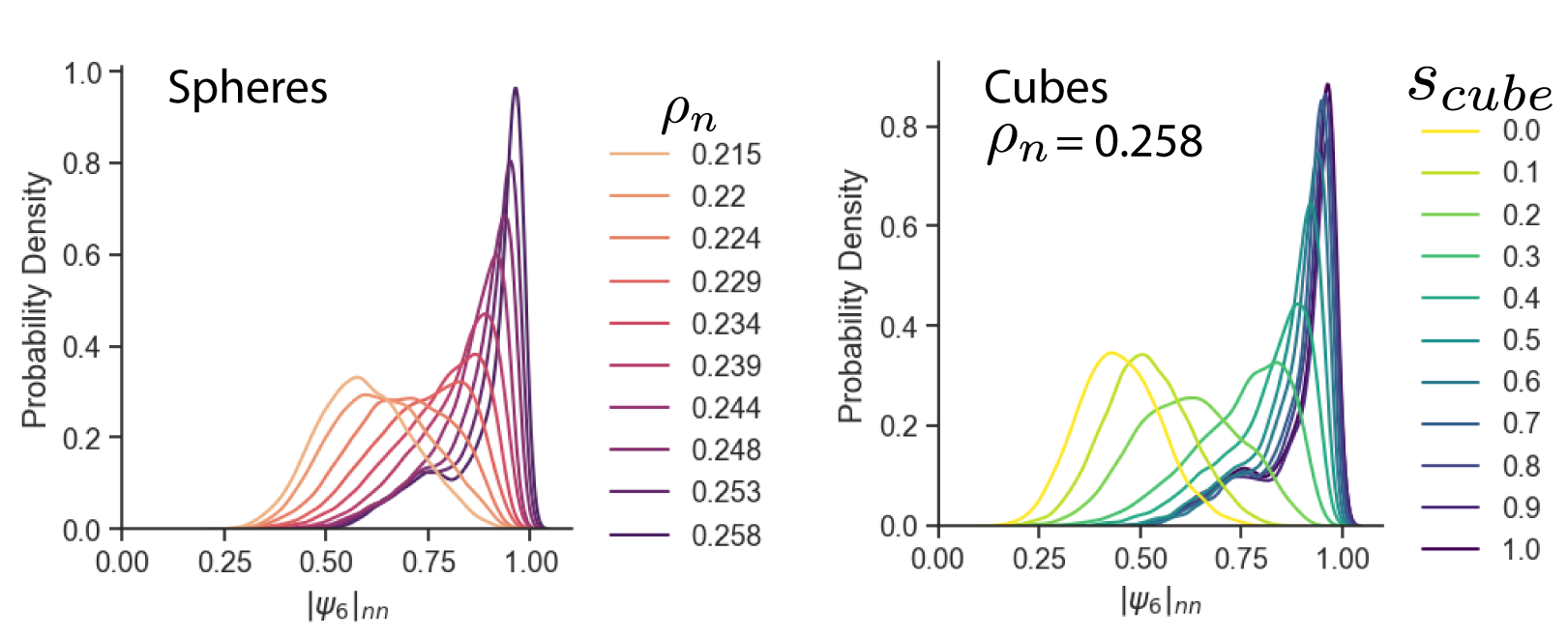}
  \caption{Distribution of $|\psi_6|_{nn}$: For systems of $n=1000$ spheres as the number density $\rho_n$ increases, the hexagonal ordering $|\psi_6|_{nn}$ increases. The hexagonal ordering also increases as the rounding of cubes $s$ increases at a constant number density $\rho_n = 0.258$. A clear minima at $\sim0.8$ justifies a cutoff of $|\psi_6|_{nn} \ge 0.8$ to denote hexagonal ordering.}
  \label{fig:PDE_Psi6}
\end{figure}

To describe the hexagonal motif we use the common order parameter $|\psi_6|$, described by $|\psi_6| = |\frac{1}{n}\sum_j^n e^{6i\theta_{ij}}|$. This quantifies six-fold symmetric rotational order about the particle. The cutoff for $|\psi_6|_{nn}$ used in the main text is justified by a minimum in the probability density estimate in figure \ref{fig:PDE_Psi6}.

\subsection{Description of the face-aligned motif}
The face-aligned motif describes the face-to-face alignment of cubes within the surface tangential plane. For each particle $i$ we calculate $\theta_{t, F}$ and $\overrightarrow{f_{t}}$. We then project the vector $\overrightarrow{f_{t}}$ onto the tangent plane of the sphere and term this $\overrightarrow{a_i}$. For all neighbor particles j, determined by distance cutoff defined by the RDF, we rotate the corresponding vector $\overrightarrow{a_{j}}$ into the reference of particle $i$ by parallel transport. We calculate the angle between these vectors and modify this by the angle $\pi/2$ to conform to the symmetry of the faces in the tangent plane, seen in equation \ref{eq:theta_ij}.

\begin{equation}
  \theta_{ij} = \left| \frac{<\overrightarrow{a_i}, \overrightarrow{a_j}>}{\|\overrightarrow{a_i}\|\|\overrightarrow{a_j}\|} \% \frac{\pi}{2} \right|
  \label{eq:theta_ij}
\end{equation}

\begin{figure}[h]
  \centering
  \includegraphics[width=0.8\linewidth]{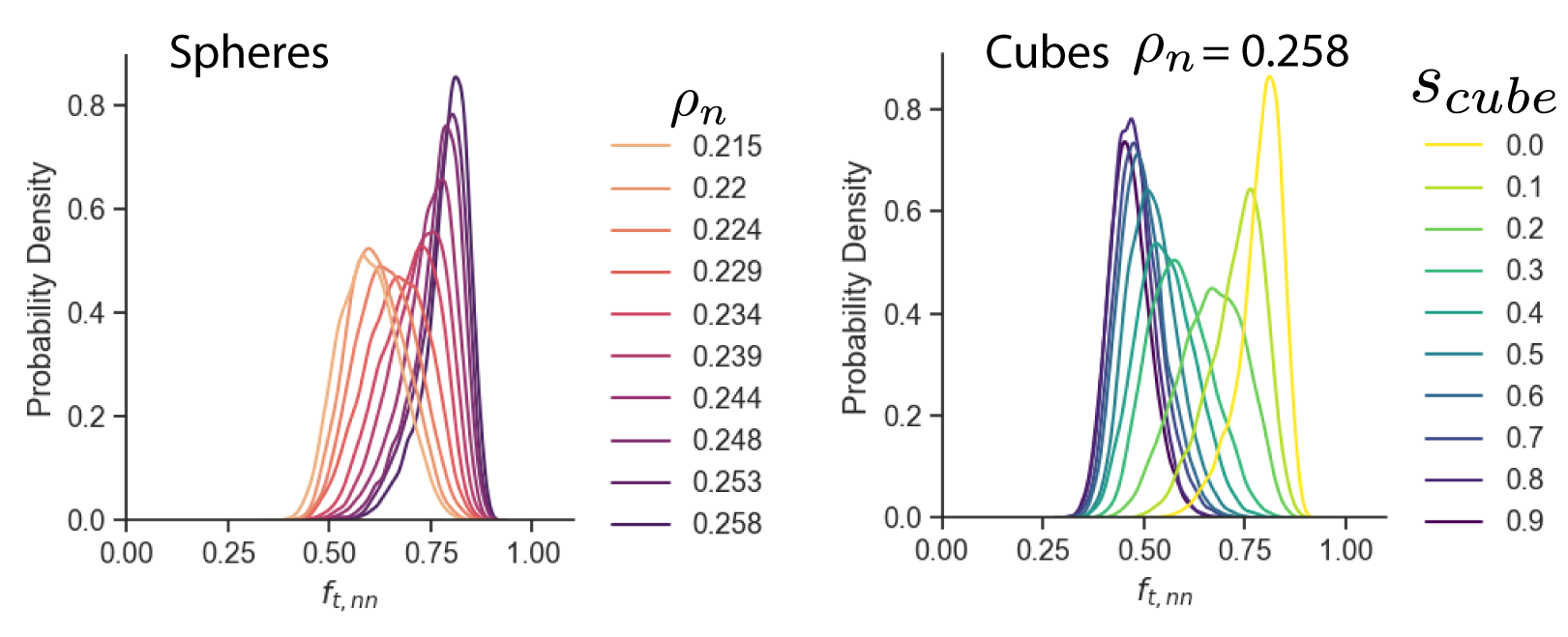}
  \caption{Distribution of $f_{nn}$: For systems of $n=1000$ rounded cubes as the number density $\rho_n$ increases or as the rounding $s$ decreases, the face-to-face alignment increases. Rounded cubes with low values of $|\psi_6|_{nn} < 0.8$ appear when values of $f_{nn}$ are greater than $\sim 0.75$, giving a reasonable cutoff for the face-aligned motif to be $f_{nn} \ge 0.75$.}
  \label{fig:PDE_Ft}
\end{figure}

Equivalent angles are visualized in figures \ref{fig:VectorAndAngleExample}b, c. We map these values of $\theta_{ij}$, over a number of neighbor particles $J$, onto the order parameter $f = (1/J)\sum_{j=1}^J\left|\frac{4}{\pi}\theta_{ij} - 1\right|$. The cutoff of $f_{nn} \ge 0.75$ is justified in figure \ref{fig:PDE_Ft}.

\subsection{Description of the woven motif}
The woven motif describes an arrangement of edge up tetrahedra, oriented such that their edges are at a $\pi/2$ angle with their four nearest neighbors. For each particle $i$ we calculate $\theta_{t, E_\parallel}$ and $\overrightarrow{e_{\parallel,t}}$. We then project the vector $\overrightarrow{e_{\parallel,t}}$ onto the tangent plane of the sphere and term this $\overrightarrow{a_i}$. For all neighbor particles j, determined by distance cutoff defined by the RDF, we rotate the corresponding vector $\overrightarrow{a_{j}}$ into the reference of particle $i$ by parallel transport. We then calculate the angle $\theta_{ij}$ between these vectors and modify this by the angle $\pi/2$ to conform to the symmetry of the woven motif, by equation \ref{eq:theta_ij}.

We map these values of $\theta_{ij}$, over a number of neighbor particles $J$, onto the order parameter $w = (1/J)\sum_{j=1}^J\left|\frac{4}{\pi}\theta_{ij} - 1\right|$.

\begin{figure}[h!]
  \centering
  \includegraphics[width=0.4\linewidth]{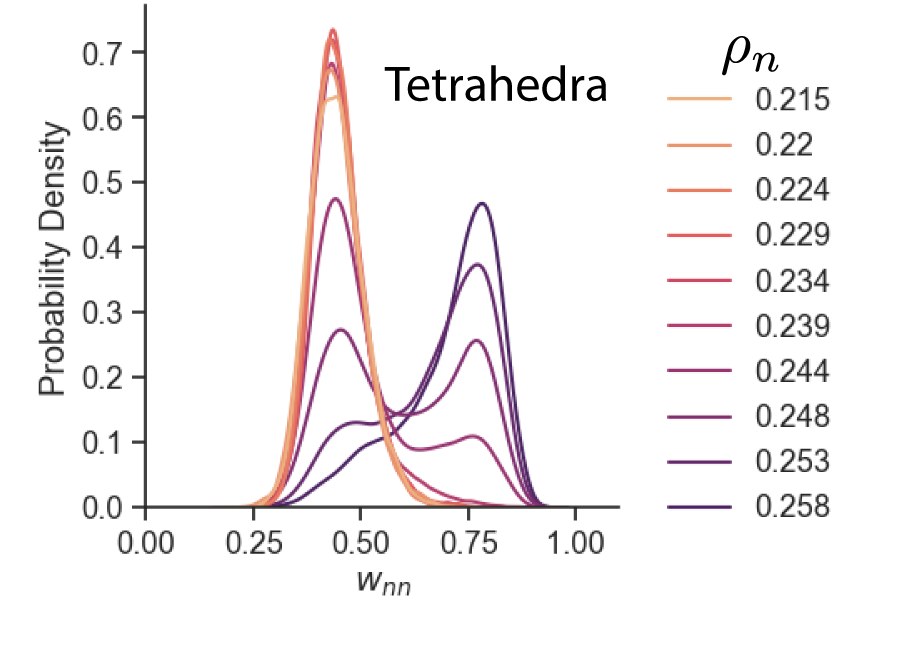}
  \caption{Distribution of $w_{nn}$: For systems of $n=1000$ tetrahedra with $s_{tet} = 0$ there is a clear shift in $w_{nn}$ at a sufficient number density $\rho_n$, and we are able to use a cutoff of $w_{nn} \ge 0.65$ that corresponds to the minima between the two peaks in the above probability density distribution to describe the woven motif.}
  \label{fig:PDE_Tetrahedron_Woven}
\end{figure}

\subsection{Description of the honeycomb motif}
The honeycomb motif describes a honeycomb structure that delineates its two basis positions by orientation, face-up and vertex-up respectively. For each particle we determine if it is face-up, vertex-up, or edge-up. We then compare the neighbors of the face and vertex up particles and if their three closest neighbors are of the opposite type (vertex and face up respectively) we calculate $hc = |\psi_3| = |\frac{1}{n}\sum_j^ne^{3i\theta_{ij}}|$. All other ineligible particles have a value of $hc=0$. 

\begin{figure}
  \centering
  \includegraphics[width=\linewidth]{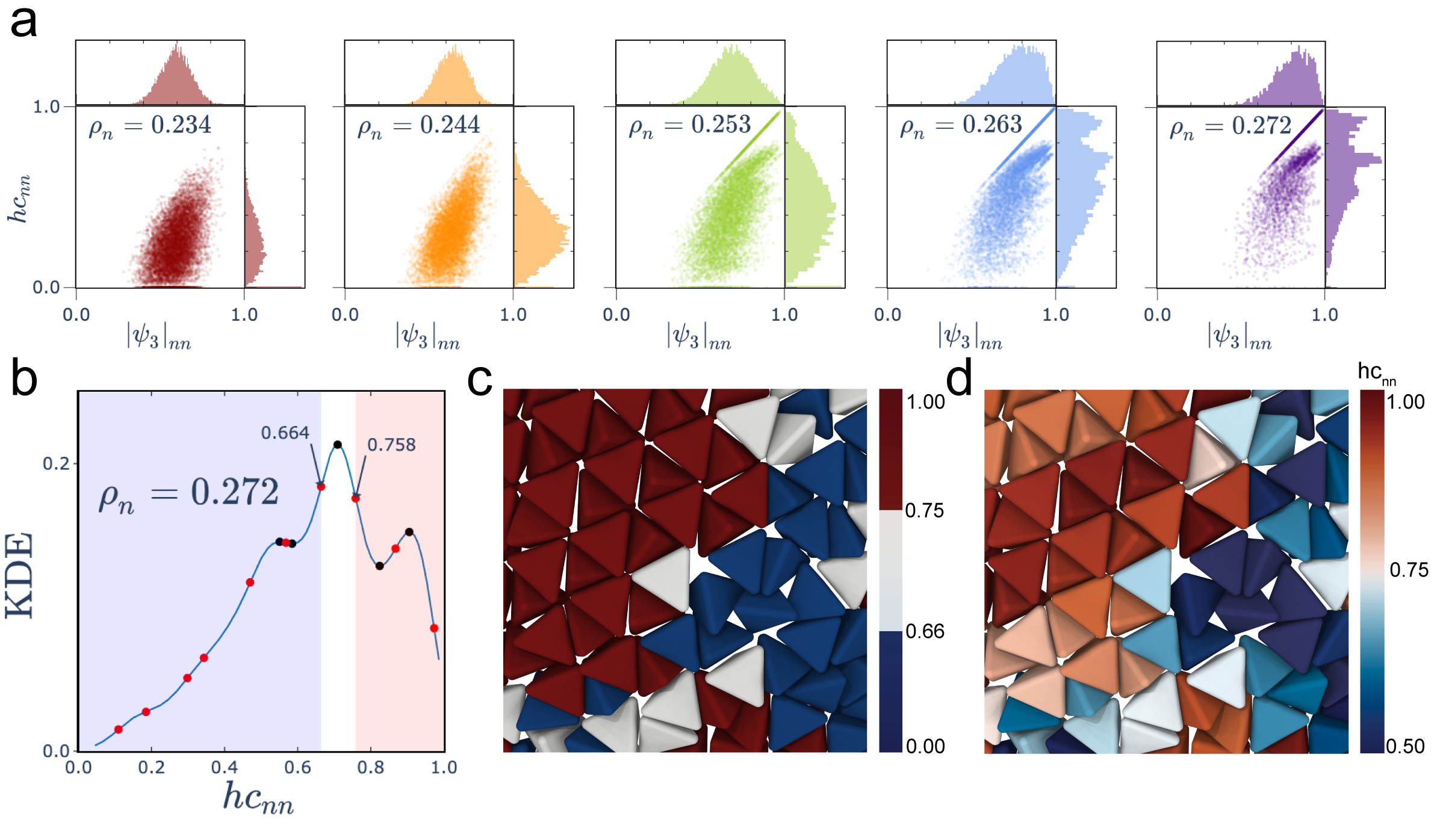}
  \caption{Distribution of $hc_{nn}$: For systems of $n=1000$ tetrahedra we considered calculating $|\psi_3|$, but this method gives no easily discernible cutoff as seen in scatter plots of $|\psi_3|_{nn} \textrm{vs} hc_{nn}$ (a). If an arbitrary cutoff of $|\psi_3|_{nn}$ is chosen by eye it often overestimates the honeycomb surface coverage, leading us to opt for the more accurate $hc_{nn}$ that takes orientation into consideration. All systems where the dominant motif is determined to be honeycomb have a characteristic line in the scatter distributions within at $|\psi_3|_{nn} = hc_{nn}$, seen in subfigure (a) $\rho_n \ge 0.253$. Inflection points I1 (left) and I2 (right) were calculated for all distributions showing this characteristic signature and showed consistent values. The first highlighted inflection point averaged to  $\textrm{I1} = 0.665 \pm 0.008$ and the second point I2 is at $\textrm{I2} = 0.756 \pm 0.007$. Given these values we use a \textbf{cutoff of $hc_{nn} \ge 0.75$}. Systems presented here are rounded by the parameter $s_{tet} = 0.2$.}
  \label{fig:HoneycombJustification}
\end{figure}

\clearpage
\subsection{System number fraction of motifs determined by relevant order parameters}

\begin{figure}[h!]
  \centering
  \includegraphics[width=\linewidth]{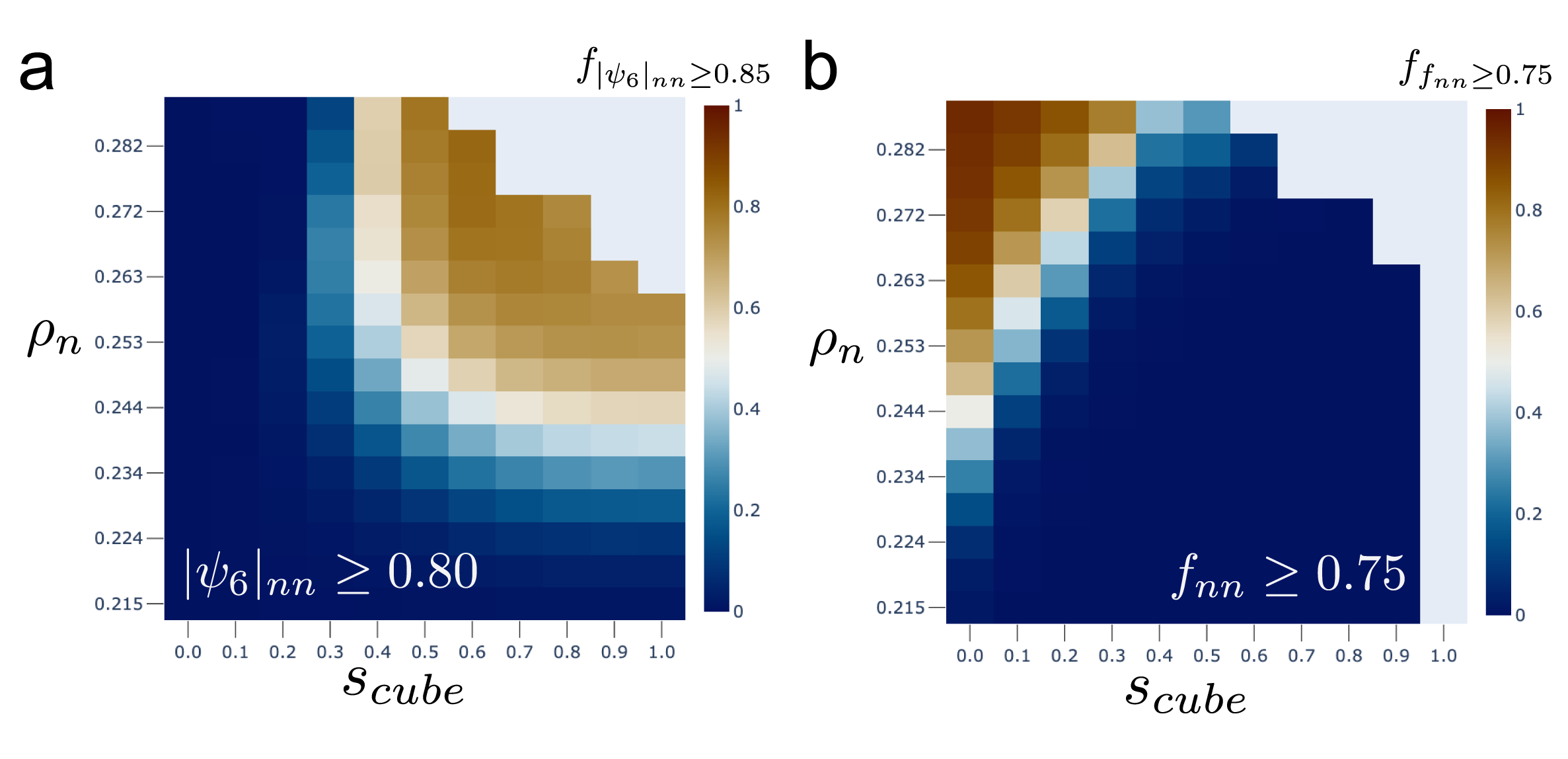}
  \caption{Both figure's text inserts correspond to the relevant order parameter for rounded cube systems. (a) System coverage of the hexagonal motif. The number of hexagonal motif particles,  $N_{|\psi_6|_{nn}\ge0.8}$, is determined using the order parameter $|\psi_6|_{nn}$ for $n=1000$ rounded cubes. The total fraction $f_{|\psi_6|_{nn}\ge0.8}=\left<N_{|\psi_6|_{nn}\ge0.8}/N_{total}\right>$, with the average taken over 10 replicates. (b) System coverage of the face-aligned motif. The number of face-aligned particles, $N_{f_{nn}\ge0.75}$, is determined using the order parameter $f_{nn}$ for $n=1000$ rounded cubes. The total fraction $f_{f_{nn}\ge0.75}=\left<N_{f_{nn}\ge0.75}/N_{total}\right>$, with the average taken over 10 replicates.}
  \label{fig:CubeCoverages}
\end{figure}

\begin{figure}[h!]
  \centering
  \includegraphics[width=\linewidth]{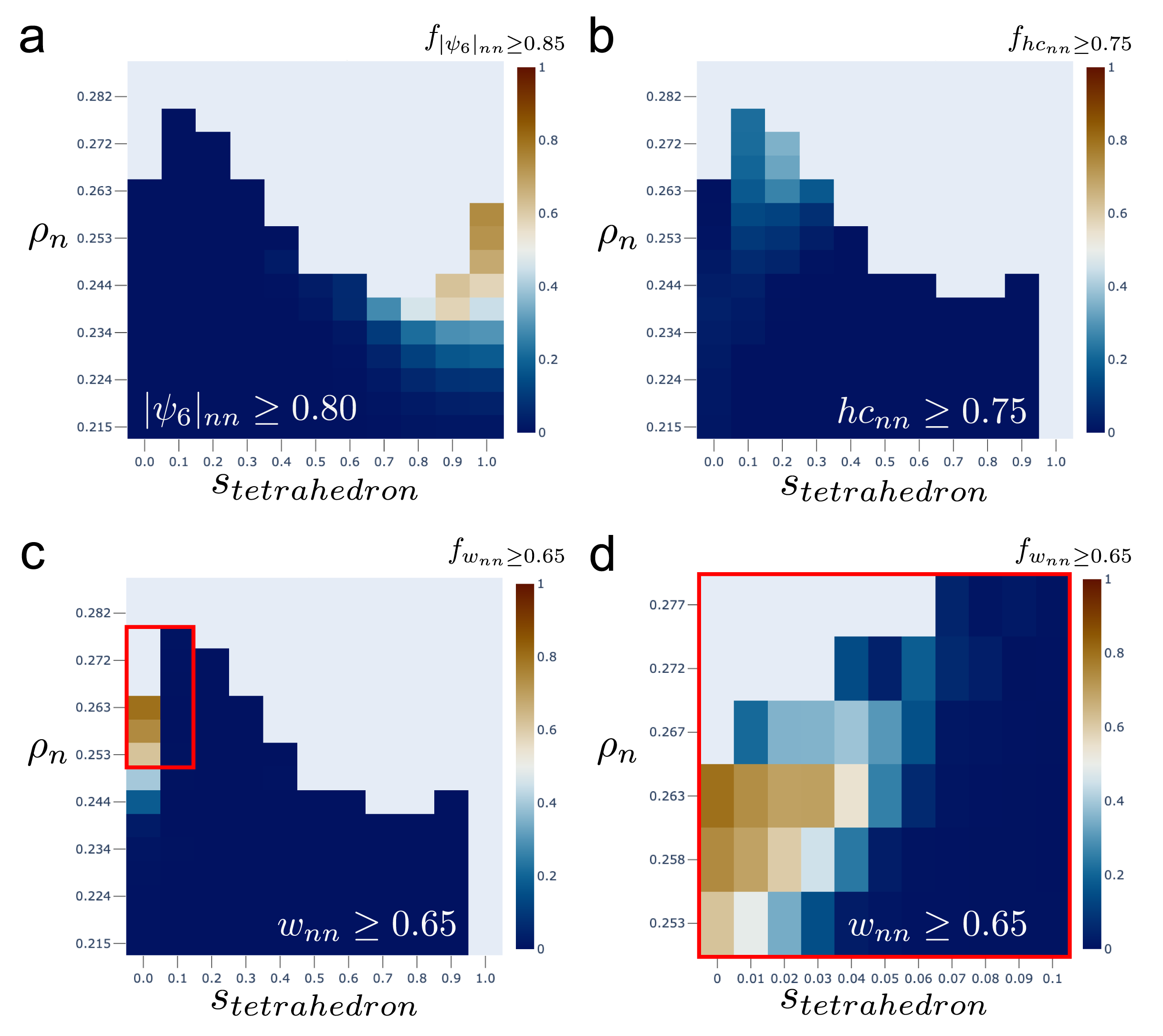}
  \caption{All figure's text inserts correspond to the relevant order parameter for rounded tetrahedron systems. (a) System coverage of the hexagonal motif. The number of hexagonal motif particles,  $N_{|\psi_6|_{nn}\ge0.8}$, is determined using the order parameter $|\psi_6|_{nn}$ for $n=1000$ rounded tetrahedra. The total fraction $f_{|\psi_6|_{nn}\ge0.8}=\left<N_{|\psi_6|_{nn}\ge0.8}/N_{total}\right>$, with the average taken over 10 replicates. (b) System coverage of the honeycomb motif. The number of honeycomb particles, $N_{hc_{nn}\ge0.75}$, is determined using the order parameter $hc_{nn}$ for $n=1000$ rounded tetrahedra. The total fraction $f_{hc_{nn}\ge0.75}=\left<N_{hc_{nn}\ge0.75}/N_{total}\right>$, with the average taken over 10 replicates. (c, d) System coverage of the woven motif. The red highlighted portion of each subfigure highlights serves to highlight the equivalent region sampled. The number of woven particles, $N_{w_{nn}\ge0.65}$, is determined using the order parameter $w_{nn}$ for $n=1000$ rounded tetrahedra. The total fraction $f_{w_{nn}\ge0.65}=\left<N_{w_{nn}\ge0.65}/N_{total}\right>$, with the average taken over 10 replicates.}
  \label{fig:TetrahedronCoverages}
\end{figure}

\clearpage
\section{S3: Defect Region Analysis}

\begin{figure}[h]
  \centering
  \includegraphics[width=\linewidth]{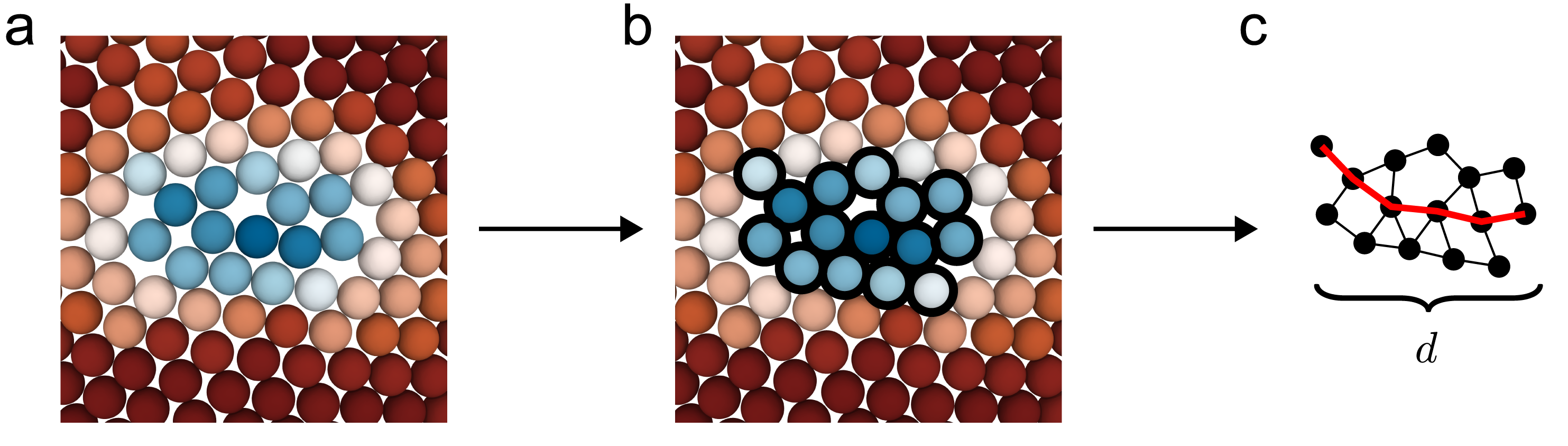}
  \caption{Illustration of longest shortest path for a collection of a particles transformed into a graph. (a) A representative image of spherical particles. (b) We consider the highlighted particles in black rings to be the defect. (c) The nodes and edges of the graph are created using the particle centroid positions and a distance cutoff. The longest shortest path of the defect determines the value of $d[\sigma]$, defect scar length.}
  \label{fig:SI_LongestShortestDefect}
\end{figure}

\begin{figure}[h]
  \centering
  \includegraphics[width=\linewidth]{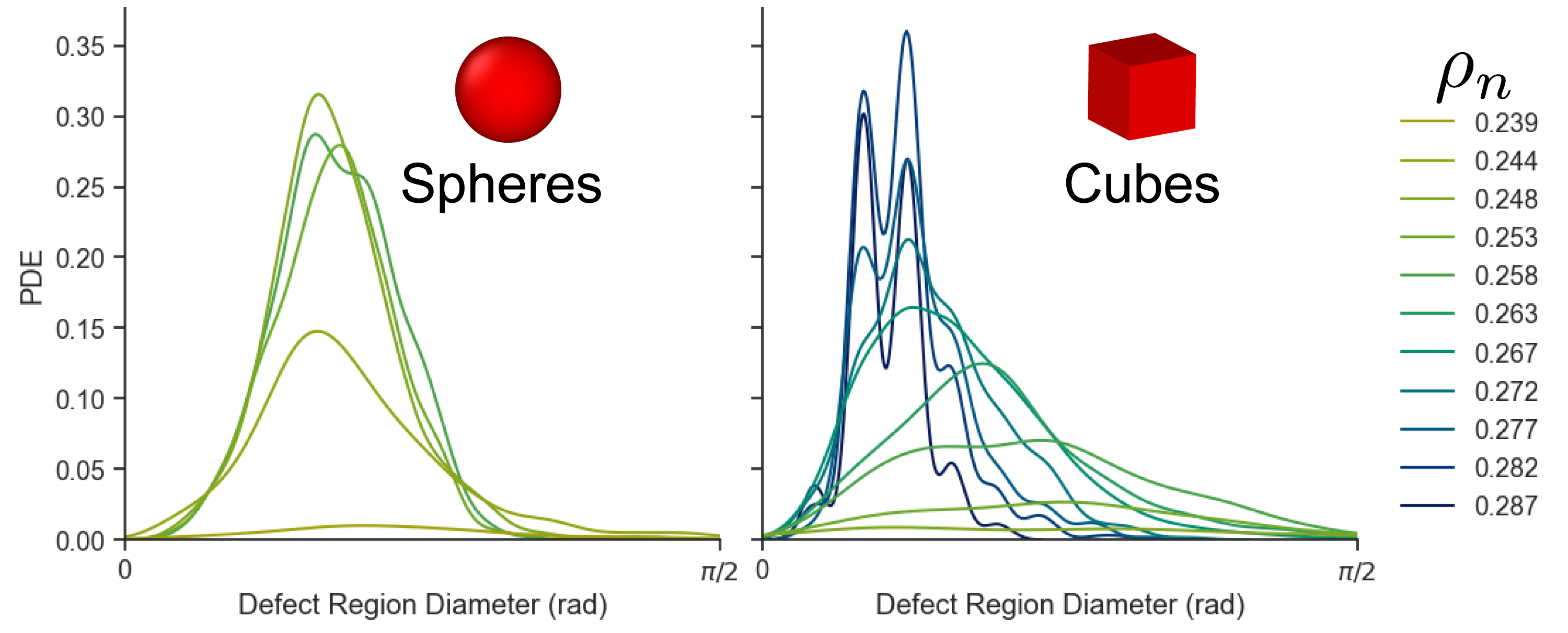}
  \caption{Scar Length PDE for spheres (left) and cubes (right) for $N=1000$ particles. Insets correspond to defects taken from the region of the PDE.}
  \label{fig:ScarLength1000PDE}
\end{figure}

Defect analysis was conducted by creating a weighted graph of particles deemed to be defects. Defects are classified either through a Voronoi tessellation or by an order parameter cutoff. These particles were then transformed to be the node positions of a graph, with edges between them populated using a distance cutoff corresponding to the first peak in the replicate-averaged radial distribution function. A visual representation of the graph is seen in figure \ref{fig:SI_LongestShortestDefect}. For each simulation frame there are a $c$ components within a graph $g$ and frames were required to have the topologically expected $c=12$ or $c=8$ components for the sphere and cube systems, respectively. The longest shortest path of each graph component was then calculated and plotted as a value $d[\sigma]$. The distribution of these defects for spheres and sharp cubes is shown in figure \ref{fig:ScarLength1000PDE}. The two distinct defect lengths for cubes are shown, within the inset.

\begin{figure}[h]
  \centering
  \includegraphics[width=\linewidth]{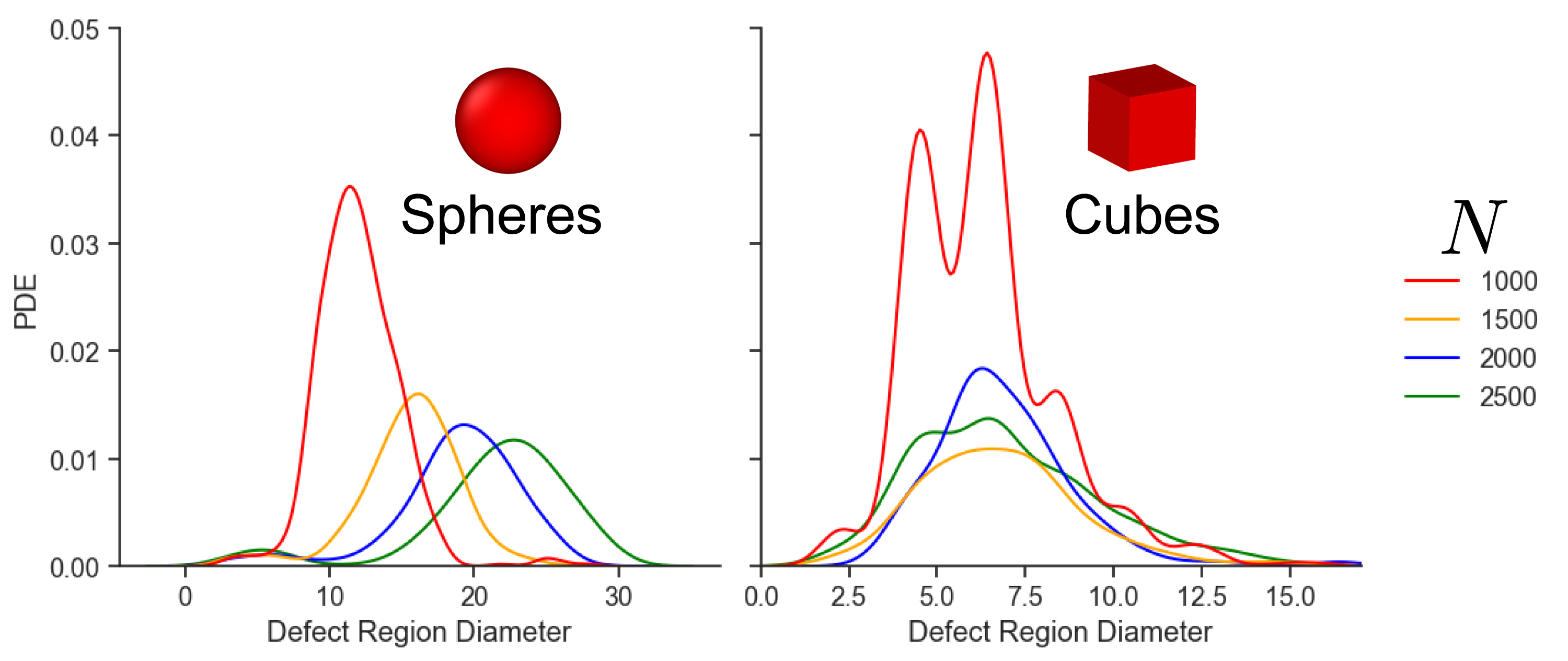}
  \caption{Scar Length PDE for spheres (left) and cubes (right) over a range of particle number N. Sphere lengths are taken with parameters $\rho_n = 0.253$ and $s=1$. Cube lengths are taken with parameters $\rho_n=0.282$ and $s=0$. Sphere scars elongate as $N$ increases, while cube defect regions shift to include only single particle guest defects. Lengths are in absolute units, not radial.}
  \label{fig:ScarLengthKDEOverN}
\end{figure}


\begin{figure*}
  \centering
  \includegraphics[width=0.7\linewidth]{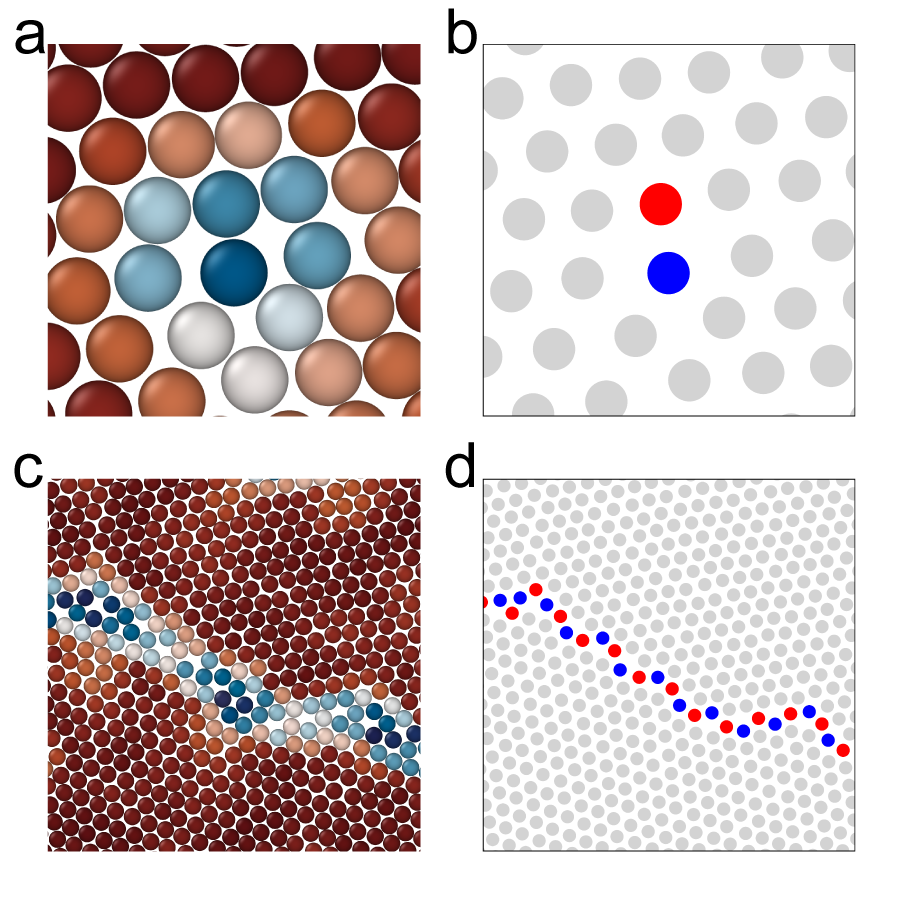}
  \caption{Visualization of a single 5-7 defect in a system of spheres (a,b). (a) The system is colored by $|\psi_6|_{nn}$. (b) A $q=+1$ defect with five-fold rotational symmetry is seen at the top in red, while a single $q=-1$ defect seven-fold rotational symmetry is seen at the bottom in blue. Visualization of a longer defect scar(c,d). (c) A system is colored by $|\psi_6|_{nn}$. (d) The same defect scar is colored by the same scheme seen in (b).}
  \label{fig:CubeDefectVisualization}
\end{figure*}

\begin{figure*}
  \centering
  \includegraphics[width=\linewidth]{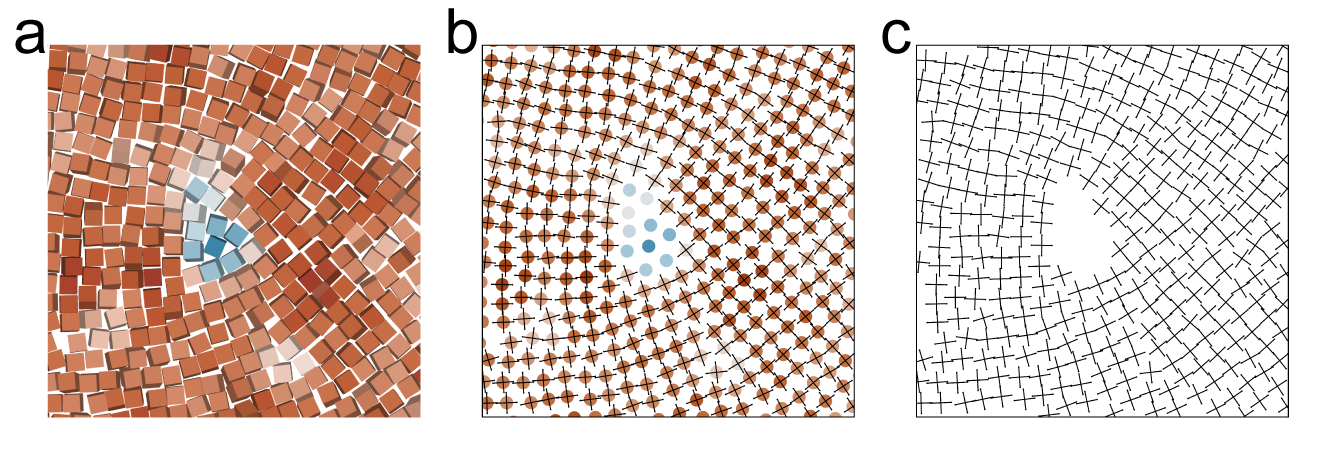}
  \caption{Visualization of a single three-fold symmetric defect in a system of cubes, with $s_{cube} = 0$. (a) Shows the system colored by $f_{nn}$. (b) shows the same system as a point cloud with tetragonal crosses overlaid which serve to guide the eye toward the three-fold morphology of the defect.}
  \label{fig:CubeDefectVisualization}
\end{figure*}

\clearpage
\newpage
\section{S4: Distributions of Motifs over a Range of N}
\begin{figure}[h!]
  \centering
  \includegraphics[width=\linewidth]{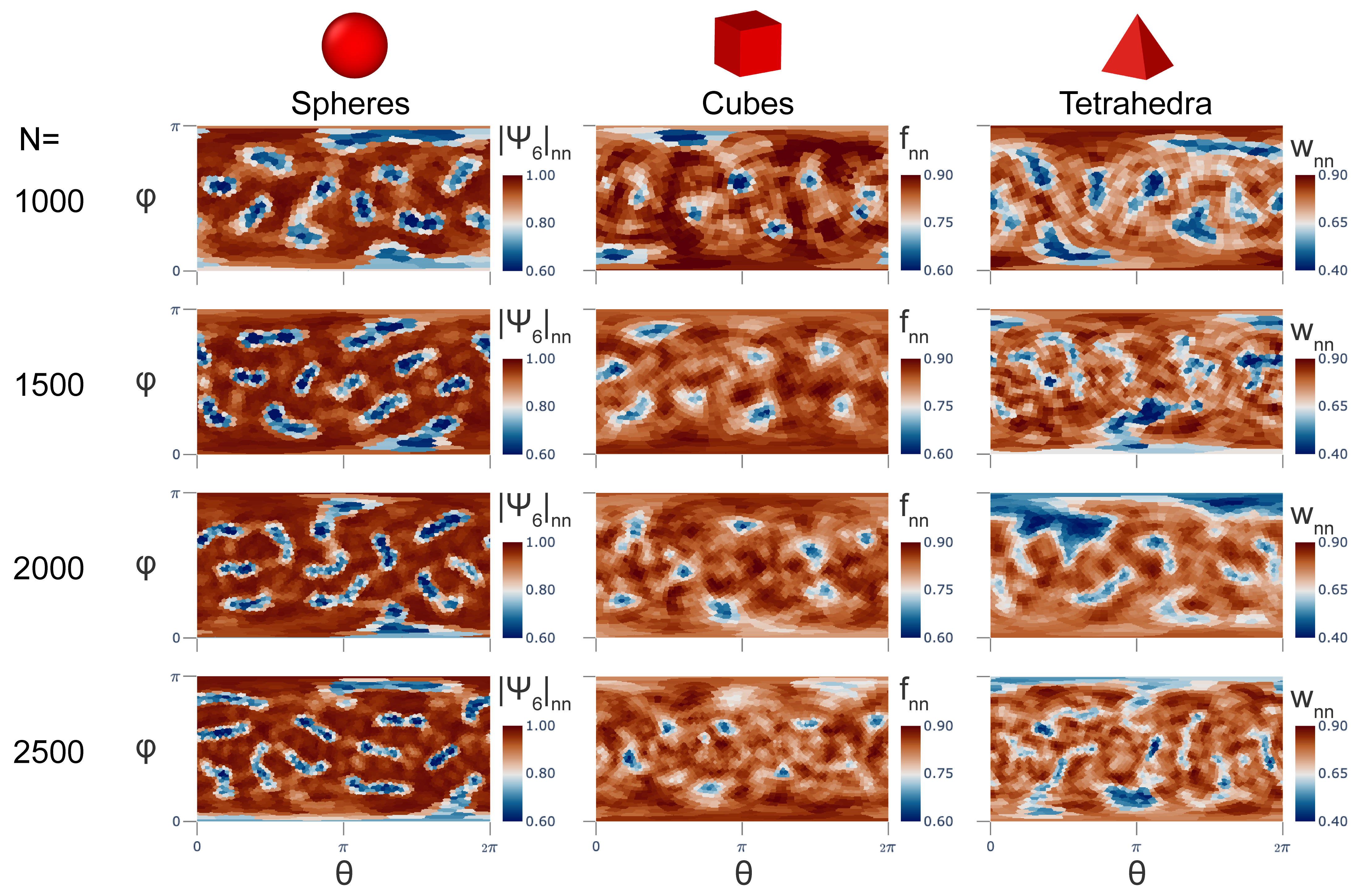}
  \caption{Illustration of mercator plots over N for each shape. (Top) spheres, with $\rho_n = 0.258$, colored by $|\psi_6|_{nn}$. (Middle) Cubes, with $\rho_n = 0.272$, colored by $f_{nn}$. (Bottom) Tetrahedra, with $\rho_n = 0.253$, colored by $w_{nn}$. These are shown across a range of $\rho_n$ values to best demonstrate the change in motif proliferation and defects across a range of $n$.}
  \label{fig:MercatorsOverN}
\end{figure}

As the number of particles $n$ increases we are able to visually see how the motif spans across the surface for each of the main motifs, hexagonal, face-aligned, and woven in figure \ref{fig:MercatorsOverN}. It should also be noted that the woven motif appeared for rounded tetrahedra on the plane, but sufficient statistics for this observation is not within the scope of this article.

%
%
%
%
%
%
%
%